\documentclass[preprint,aps,pra,showpacs,epsfig,preprintnumbers,superscriptaddress,floatfix]{revtex4}
\usepackage{bm}
\usepackage{latexsym}
\usepackage{epsfig}
\usepackage{graphicx}
\usepackage{textcomp}
\usepackage{graphicx}
\usepackage{dcolumn}
\usepackage{subfigure}
\usepackage{subeqn}

\bibstyle{prsty}

\begin{document}

\title{Spontaneous creation and persistence of ground-state coherence in a resonantly driven intra-cavity atomic ensemble}
\author{D. G. Norris}
\affiliation{ Joint Quantum Institute, Department of Physics, University of Maryland and National Institute of Standards and Technology, College Park, MD 20742 USA}
\author{A. D. Cimmarusti}
\affiliation{ Joint Quantum Institute, Department of Physics, University of Maryland and National Institute of Standards and Technology, College Park, MD 20742 USA}
\author{L. A. Orozco }
\affiliation{ Joint Quantum Institute, Department of Physics, University of Maryland and National Institute of Standards and Technology, College Park, MD 20742 USA}
\author{P. Barberis-Blostein}
\affiliation{Instituto de Investigaci{\'o}n en Matem{\'a}ticas Aplicadas y en Sistemas, Universidad Nacional Aut{\'o}noma de M{\'e}xico, M{\'e}xico, DF 01000, M{\'e}xico}
\author{H. J. Carmichael}
\affiliation{Department of Physics, University of Auckland, Private Bag 92019, Auckland, New Zealand}

\date{\today}

\begin{abstract}
The spontaneous creation and persistence of ground-state coherence in an ensemble of intracavity Rb atoms has been observed as a quantum beat. Our system realizes a quantum eraser, where the detection of a first photon prepares a superposition of ground-state Zeeman sublevels, while detection of a second erases the stored information. Beats appear in the time-delayed photon-photon coincidence rate (intensity correlation function). We study the beats theoretically and experimentally as a function of system parameters, and find them remarkably robust against perturbations such as spontaneous emission. Although beats arise most simply through single-atom-mediated quantum interference, scattering pathways involving pairs of atoms interfere also in our intracavity experiment. We present a detailed model which identifies all sources of interference and accounts for experimental realities such as imperfect pre-pumping of the atomic beam, cavity birefringence, and the transit of atoms across the cavity mode.
\end{abstract}


\maketitle
\section{Introduction}
\label{sc:introduction}
The interference of scattering amplitudes in quantum mechanics arises from the indistinguishability of alternative scattering paths, as in the interference of the paths of a photon passing through the celebrated double-slits of Young. For example, a modern variation on the Young experiment \cite{eichmann93} shows spatial fringes in the intensity of light scattered from two trapped ions; but only so long as the scattering cannot be traced to one ion or the other \cite{itano98}.  Even in the latter situation, other measurements might be found which cannot distinguish between the paths, hence recovering the interference. Typically they involve higher-order moments of the field \cite{agarwal02}, or post-selection, \emph{i.e.}, partitioning of scattering events into subensembles \cite{scully82}.  Such a measurement strategy is termed a ``quantum eraser'', since it recovers interference by ``erasing'' the information that identifies the path.

A time-domain analogue of double-slit interference can occur inside multi-level atoms, where photon emission via parallel
transitions can result in a modulation of the emission intensity at the frequency of a level splitting---the phenomenon of ``quantum beats" (see, \emph{e.g.}, \cite{aleksandrov64,dodd67}).  A distinction has traditionally been drawn between ``Type-I'' or ``V'' atomic systems, where decay of a superposition of upper levels yields beats at the transition difference frequency; and the inverted ``Type-II'' or ``$\Lambda$'' systems, where decay to a superposition of lower levels \emph{does not} yield beats \cite{breit33,ficek05book}.  The typical argument for the latter outcome is that a measurement of the ground-state population could always, in principle, determine in which of the two available states the electron landed; as there is no sum over alternative paths to one and the same final state, there are no ``ground-state quantum beats'' \cite{chow75,herman75}.

Nevertheless, as with spatial fringes, a quantum eraser-type strategy can recover time-domain interference in the ground state.  Zajonc \cite{zajonc83,ficek05book} proposed one such implementation in a ``Type-II'' atomic system, basing his proposal on two-photon scattering. In this case the second scattered photon erases the path information written by the first---amplitudes for the scattering of two photons in sequence interfere.

We recently published experimental results showing quantum beats in spontaneous emission at the frequency of the ground-state Zeeman splitting in Rb, $\textit{i.e.}$ ground-state quantum beats seen in spontaneous emission \cite{norris10}. Oscillations appear in the second-order intensity autocorrelation function only, not in the average intensity, as follows from the indistinguishability requirement above.  The presence of a similar oscillation hidden within the noise of spontaneous emission was demonstrated in 1955 by Forrester \emph{et al.}~\cite{forrester55}. They mixed two incoherent light sources---a Zeeman doublet---on a photocathode, and used a resonant microwave cavity to enhance the beat signal extracted from the photocurrent noise.  The interference in this case is classical, though the oscillation is recovered from noise through intensity correlation. More than 50 years later, using coherent excitation and single-photon detectors, we have realized a time-resolved measurement of the ground-state quantum beat recovered from spontaneous emission noise.

The oscillation in our system arises as a complicated mixture of quantum eraser-type interferences within single atoms, pair-wise interference between emission from different atoms, and a homodyne contribution due to the superposition of a weak coherent
background (similar to \cite{gerber09}) generated by birefringence in the cavity mirrors. By coupling spontaneous emission into an optical cavity at moderate dipole coupling strength, we overcome the signal-to-noise limitations set by a small coherence area in free space \cite{forrester55}, and enforce indistinguishability among different atoms emitting into a common spatial mode.  Moreover, we show below that the complicated level structure of $^{85}$Rb actually aids in the survival of ground-state coherence, counter to the conventional strategy of protecting coherence by limiting the state space through which population can diffuse.

We distinguish here between ground-state coherences imposed by an external drive and those arising spontaneously, selected through the detection process, as in our experiment.  In the former case, an external magnetic or optical drive couples two ground states directly, with the resulting coherence read out optically in forward scattering (see \cite{alexandrovbook} for many examples.) In the latter, levels couple only through the vacuum, with no external drive to enforce coherence.  The fact that spontaneous decay can generate coherence is evident from the observation of quantum beats at the intermediate level splitting in cascade decay \cite{aspect84,willis10}; that the same process occurs in transitions to ground or meta-stable states is not therefore surprising. Schubert et al. \cite{schubert95} measured such a coherence in the bichromatic cross-correlation of fluorescence from a single ion, where detection of a first photon left the ion in a superposition of meta-stable states.  When considering isotropic emission, however, spontaneously created coherences tend to vanish on the average, and for this reason are often left out of density matrix calculations \cite{aspect89}. The recent interest follows a 1992 paper \cite{javanainen92} in which specific measurable consequences were claimed; various arrangements have been explored theoretically \cite{patnaik99,kiffner06}.  A publication as recent as 2005  \cite{dutt05} claims evidence of the first serious experimental consequences of spontaneously-generated coherence, this in a quantum dot system.  Other experiments are surprisingly few.  We direct the interested reader to Ref.~\cite{economou05} for an overview.

In this article we expand upon the results presented in \cite{norris10}. In particular, we seek to explain the origin of the various individual components of the beat signal, show which experimental conditions are necessary for the robust survival and detection of beats, and explore their sensitivity to various experimental controls.  The paper is organized as follows. Section \ref{sc:theoreticalmodel} introduces the theoretical model, starting with a single atom fixed in space and  moving to a full atomic beam with realistic fluctuations. Section \ref{sc:experiment} presents the details of our experimental method, from the atomic source and optics to the detection apparatus.  Section \ref{sc:results} summarizes the evolution of the beat signal as we explore parameter space and compares experiment with theory.  The paper concludes in Section \ref{sc:conclusions}.

\section{Theoretical model}
\label{sc:theoreticalmodel}

\begin{figure}
\begin{center}
\includegraphics[width=0.90\linewidth]{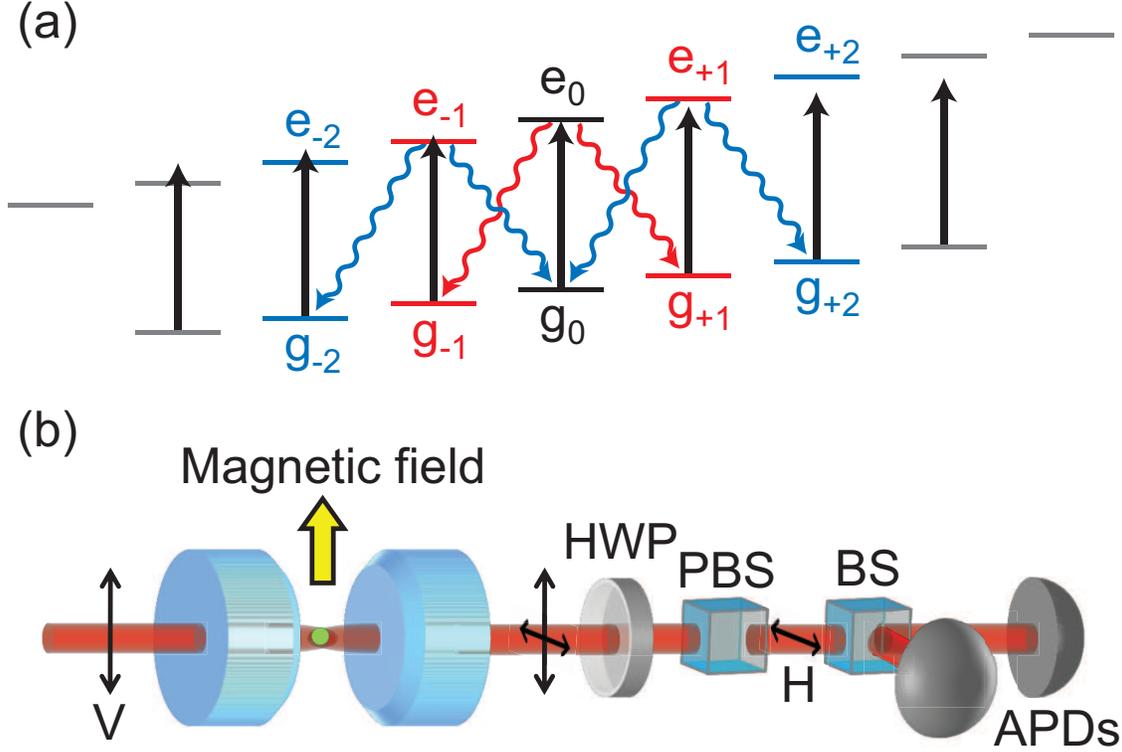}
\caption{\label{levels}
(Color online) Experimental setup: (a) $\pi$-excitation of the $F=3$ to $F^\prime=4$ transition in Rb showing scattering of a first (red) and second (blue) photon into the $H$ mode; (b) schematic of the apparatus, HWP: half-wave plate, PBS: polarizing beam-splitter, BS: beam-splitter, APD: avalanche photodiode.}
\end{center}
\end{figure}

\begin{figure}
\begin{center}
\includegraphics[width=0.90\linewidth]{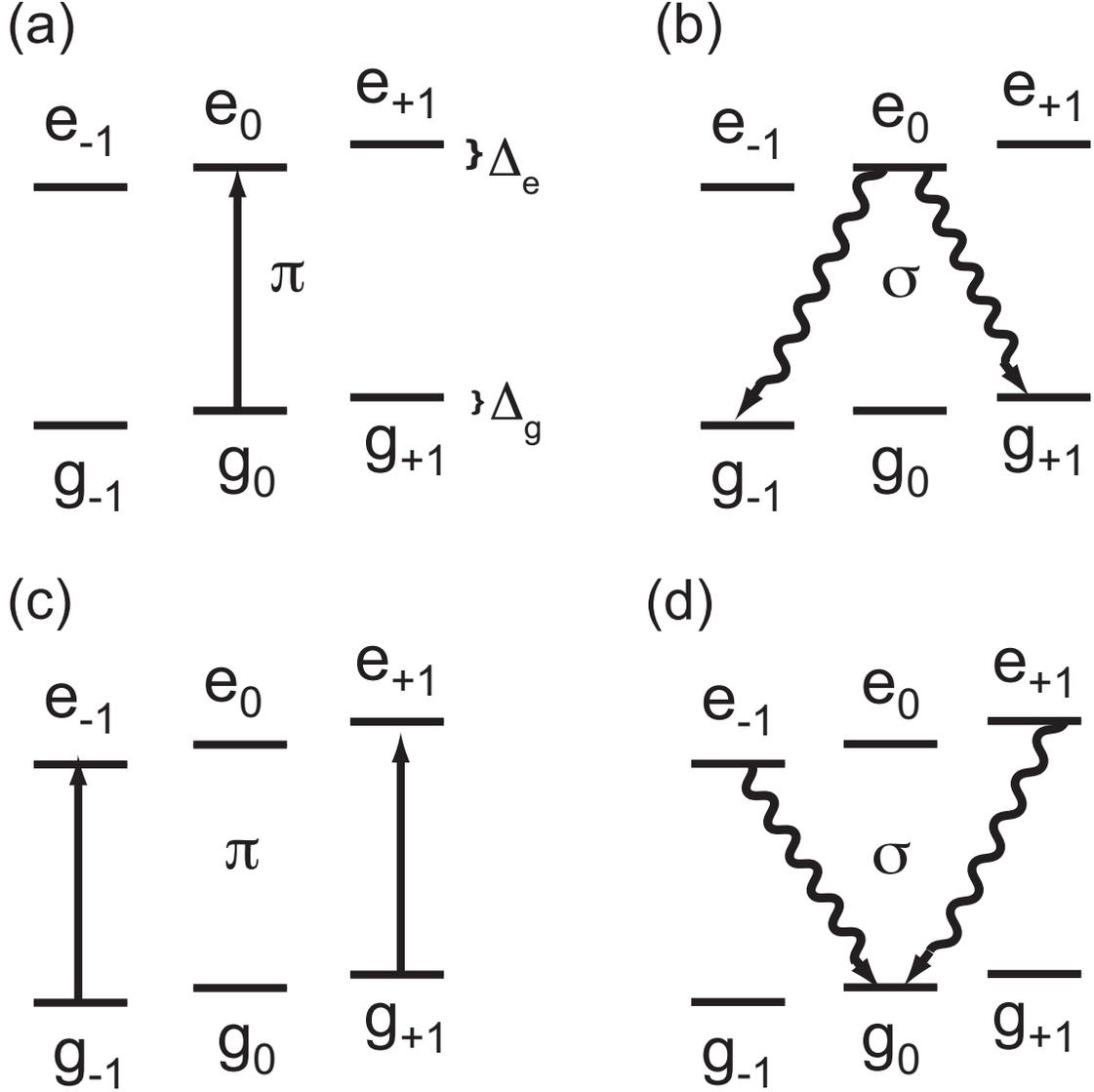}
\caption{\label{eraser} Simplified model of the two-photon quantum eraser process. (a) $\pi$-excitation from $g_0$ to $e_0$; (b) spontaneous decay through $\sigma$ transitions to a superposition of $g_{-1}$ and $g_{+1}$; (c) $\pi$-excitation to a superposition of $e_{-1}$ and $e_{+1}$; (d) spontaneous decay through $\sigma$ transitions back to $g_0$.}
\end{center}
\end{figure}

We consider first the idealized system of one fixed $^{85}$Rb atom, then turn to a realistic atomic ensemble, as realized in our experiment with a cold atomic beam.  The atom has Zeeman structure in its ground and excited states [Fig.~\ref{levels}(a)] and interacts through the $D_2$ line with degenerate, orthogonally polarized cavity modes, designated $H$ (horizontal) and $V$ (vertical); a weak magnetic field sets the quantization axis in the vertical direction, and mode $V$ is weakly and continuously driven [Fig.~\ref{levels}(b)].  The atom is prepared in state $|g_0\rangle$, from which it is excited to state $|e_0\rangle$ by the $V$ mode [Fig.~\ref{eraser}(a)]. It may return to the ground state by emitting a $\pi, \sigma^{+}$ or $\sigma^{-}$ photon, or any linear combination conserving angular momentum. In the assigned geometry, only $\sigma^{+}$ or $\sigma^{-}$ light couples to the $H$ mode, with the helicity undetermined.  We assume that the probability of reabsorption of an emitted photon is negligible; it escapes the cavity and its detection places the atom in the superposition [Fig.~\ref{eraser}(b)]:
\begin{equation}
\label{eq:super0ground}
|\psi_0\rangle=(|g_{-1}\rangle+|g_{+1}\rangle)/\sqrt{2}\, .
\end{equation}

The atom is now in the ground state with its angular momentum perpendicular to the magnetic field, and it performs Larmor precession. With subsequent reexcitation by the $V$ mode, the state
\begin{equation}
\label{eq:super0excited}
|\psi_0^\prime(t)\rangle=(e^{-i\phi(t)}|e_{-1}\rangle+e^{i\phi(t)}|e_{+1}\rangle)/\sqrt{2}
\label{eq:superphase}
\end{equation}
is reached, with phase $\pm\phi(t)$ gained through precession in the ground state [Fig.~\ref{eraser}(c)]. From here the atom can decay back to $|g_0\rangle$ by emitting a second $H$-mode photon [Fig.~\ref{eraser}(d)].  The probability for this emission depends on the phase difference, $2\phi(t)$, between the two parts of the superposition. It oscillates and thus gives rise to beats in the rate of detection of a second $H$ photon subsequent to the detection of a first.

As depicted in Fig.~\ref{eraser}, there are two paths for scattering a pair of photons into the $H$ mode: $|g_0\rangle\rightarrow|e_0\rangle\rightarrow|g_{+1}\rangle\rightarrow|e_{+1}\rangle\rightarrow|g_0\rangle$
and $|g_0\rangle\rightarrow|e_0\rangle\rightarrow|g_{-1}\rangle\rightarrow|e_{-1}\rangle\rightarrow|g_0\rangle$. The
phase gained from the ground-state Zeeman shift (Larmor precession) is different along the two paths, which interfere to produce oscillations in the rate of delayed coincidences---\textit{i.e.}, in the correlation function $g^{(2)}(\tau)$. Note that after the first photon is detected, ``which path'' information is available, since $|g_{+1}\rangle$ and $|g_{-1}\rangle$ are distinguishable in principle, and their orthogonality precludes observation of interference effects in the average intensity arising from cross-terms when taking the expectation value over Eq.~\ref{eq:super0ground}.  This information is largely erased by the second photon detection, where the amplitude for returning to the common final state $|g_{0}\rangle$ allows survival of cross-terms in the two-photon intensity expectation value, corresponding to interference between the two scattering paths.  We note that, as in Ref.~\cite{forrester55}, the spectrum of scattered light still exhibits a doublet separated by the beat frequency, but the random phase relation between the two fields destroys any first-order coherence in the average intensity.

In the following sections we study the dependence of the quantum beats on the different features and parameters of our experimental system: magnetic field strength, number of atoms, initial state preparation, atomic beam fluctuations, cavity birefringence, and mixing of the undriven $H$ field with drive light ($V$ field) outside the cavity before detection. We first present a detailed theory for the case of one fixed atom before we move on to the treatment of many atoms and atomic motion.

\subsection{One fixed atom}\label{sc:fixedatom}
The relevant 16-level structure of the atom is depicted in Fig.~\ref{levels}. Bold black arrows indicate its interaction with the driven $V$ mode of the optical cavity, and red and blue wavy lines with the undriven $H$ mode. We calculate the second-order correlation function of the $H$ mode. We use the quantum trajectory formalism \cite{carmichael93book}, which provides insight into the physical processes involved and facilitates efficient numerical calculations, something of importance when many atoms are considered.  Working in a frame rotating at the frequency of the drive, the non-Hermitian Hamiltonian governing coherent evolution between spontaneous emission events (modes other than $H$ and $V$) or photon loss through the cavity mirrors is:
\begin{equation}
\label{eq:hamiltonian}
H_S=H_0+H_I+H_D+H_L\, ,
\end{equation}
with free Hamiltonian
\begin{subequations}
\begin{eqnarray}
\label{eq:hamiltonian_free}
H_0&=&\hbar\delta_v a_v^\dagger a_v+\hbar\delta_h a_h^\dagger a_h\nonumber\\
&&+\sum_{i=-3}^3 \hbar\delta_{g_i}|g_i\rangle\langle g_i|+\sum_{i=-4}^4\hbar\delta_{e_i}|e_i\rangle\langle e_i|\, ,
\end{eqnarray}
interaction and drive
\begin{eqnarray}
\label{eq:hamiltonian_interaction}
H_I&=&\hbar g[a_v^\dagger\Sigma_\pi+a_h^\dagger(\Sigma_{\sigma_+}+\Sigma_{\sigma_-})]+{\rm h.c.}\, ,\\
\label{eq:hamiltonian_drive}
H_D&=&i\hbar\mathcal E(a_v^\dagger-a_v)\, ,
\end{eqnarray}
and non-Hermitian loss term
\begin{eqnarray}
\label{eq:hamiltonian_loss}
H_L&=&-i\hbar\kappa(a_v^\dagger a_v+a_h^\dagger a_h)\nonumber\\&&
-i\hbar\frac{\gamma}{2}(\Sigma^{\dagger}_\pi\Sigma_\pi+\Sigma^{\dagger}_{\sigma_+}\Sigma_{\sigma_+}+
\Sigma^{\dagger}_{\sigma_-}\Sigma_{\sigma_-})\, ,
\end{eqnarray}
\end{subequations}
where $a_v$ and $a_h$ annihilate photons in the $V$ and $H$ modes, respectively, detunings from the drive $\delta_v$ and $\delta_h$, $\mathcal{E}$ is the drive amplitude for the $V$ mode, $g$ is the dipole coupling constant, $2\kappa$ is the photon loss rate from each cavity mode, $\gamma$ is the spontaneous emission rate, and $\Sigma_{\pi}$, $\Sigma_{\sigma_+}$, and $\Sigma_{\sigma_-}$  are dipole lowering operators for $\pi$, $\sigma_+$, and $\sigma_-$ transitions; atomic energy shifts in the magnetic field and relative to the drive are $\hbar\delta_{e_i}$, $i=-4,-3,\ldots,4$ (excited states) and $\hbar\delta_{g_i}$, $i=-3,-2,\ldots,3$ (ground states). Explicit forms for the dipole operators  depend on Clebsh-Gordon coefficients and are relegated to the Appendix.

The evolution of the system under the non-Hermitian Hamiltonian $H_S$ is calculated numerically, with photon-number truncation, using the fourth-order Runge-Kutta method, including step-size correction. At regular intervals of length $\Delta t$ the atom and the cavity modes are checked for quantum jumps---spontaneous emission or photon leakage. If a scattering event occurs, the system state is collapsed accordingly---jump operators $\Sigma_\pi$, $\Sigma_{\sigma_+}$, $\Sigma_{\sigma_-}$, $a_v$, or $a_h$---before continuing the coherent evolution. After a time $t$, we assume a photon is scattered through the $H$ mode. The system state is collapsed (jump operator $a_h$) and its conditional evolution, with the unnormalized correlation function, $G^{(2)}(t,t+\tau)$, given by the $H$-mode-photon-number-weighted mean of the $H$-mode photon number expectation in the conditional states. Normalization by the photon number averages yields $g^{(2)}(t,t+\tau)=G^{(2)}(t,t+\tau)/\langle(a_h^\dagger a_h)(t)\rangle\langle (a_h^\dagger a_h)(t+\tau)\rangle$.

We consider now a weak drive, such that $\langle a_v^\dagger a_v\rangle\ll1$, and $t$ long enough for the system to reach a quasi-steady-state (overlooking the very slow process of optical pumping). If the initial state is chosen to be any one of the ground states $|g_i\rangle$, $i=-2,-1,0,+1,+2$, the correlation function, $g^{(2)}(\tau)$, shows beats, with beat frequency twice the Larmor precession frequency of $|g_{\pm1}\rangle$; the frequency does not depend on the chosen initial state, or the number of atoms when many are included (Sec.~\ref{sc:manyatoms}).

The visibility of the beat depends, however, on a number of issues. In the ideal case, the atom is in state $|g_0\rangle$ when the first $H$-photon is detected; it is projected to $|\psi_0\rangle$, and the second photon is detected as it returns to $|g_0\rangle$ via $|\psi^\prime_0(t)\rangle$ (Fig.~\ref{eraser}). We note first that, for the level scheme of Fig.~\ref{levels}, the return is not to $|g_0\rangle$ but (see the Appendix for the explicit form of the lowering operator, $\Sigma_{\sigma_+}+\Sigma_{\sigma_-}$, for $H$-mode scattering)
\begin{equation}
\label{eq:return}
2\cos[\phi(t)]\sqrt{\frac5{14}}|g_0\rangle+\sqrt{\frac3{28}}(e^{-i\phi(t)}|g_{-2}\rangle+e^{i\phi(t)}|g_{+2}\rangle).
\end{equation}
This brings a reduction of the visibility to a little more than 75\%. Beyond this, several trajectories deviate from the ideal and further reduce the visibility. We divide them into two groups: (i) those that deviate prior to the detection of the first photon, and (ii) those that deviate between the detection of the first photon and the second.

Consider first group (i). If the atom is initially in state $|g_i\rangle$, $i\neq 0$, then when the first $H$-photon is detected, it is projected into an unequal superposition of $|g_{i\pm1}\rangle$, with the different weights given by Clebsch-Gordan coefficients; for example, if the initial state is $|g_{+1}\rangle$, after detection of a scattered photon, the atom is projected into $|\psi\rangle=(\sqrt{10}|g_{0}\rangle+\sqrt{3}|g_2\rangle)/\sqrt{13}$. Table~\ref{tb:visibility} lists the realized visibilities for all potential initial states. A maximum visibility of $0.75$ is achieved when the initial state is $|g_0\rangle$. The visibility decays rather quickly away from this maximum. Experimentally, we optically pump the atoms before they enter the cavity; however, the efficiency of the pumping is not perfect, and the result is generally a distribution over ground states, peaked around $|g_0\rangle$. Some reduction of the beat visibility must follow from the imperfect optical pumping prior to an atom entering the cavity.
\begin{table}
\begin{tabular}{| l | c c c c|}
\hline
Initial state & $g_0$ & $g_{+1}$ & $g_{+2}$ & $g_{+3}$ \\
Visibility & 0.75 & 0.5 & 0.15 & 0.03 \\
\hline
\end{tabular}
\caption{\label{tb:visibility} Quantum beat visibility for one fixed atom and different initial ground states. The parameters are: $g/\gamma=0.25$, $\kappa/\gamma=0.5$, $\mathcal{E}/\gamma=0.025$, and a Larmor frequency of $2.2\gamma/3$. The time at which the first $H$-mode photon is detected is $t=25\gamma^{-1}$. Results for $g_{-1}$, $g_{-2}$, $g_{-3}$ follow by symmetry. Note that a visibility of zero (no quantum beat) is predicted for $g\pm3$ in the absence of optical pumping prior to $t$.}
\end{table}

Other processes can redistribute population amongst the atomic levels and contribute to the loss of visibility. The atom can spontaneously decay from an excited state $|e_i\rangle$ to $|g_{i+1}\rangle$ or $|g_{i-1}\rangle$ with the emission of a photon to the side rather than into the $H$ mode of the cavity. At higher values of the drive, this process might be repeated many times, redistributing population before a first $H$-photon is detected.  The distribution reached through such optical pumping by the drive depends on the Clebsch-Gordan coefficients, the drive strength, and the time when the first photon is detected.  Table~\ref{tb:prob} displays the distribution reached in the long-time limit.  It shows that, even in the limit, states with $i=\pm3$ have a very small probability to be populated, and those with $i=\pm2$ are populated at a level of only about 10\%.

\begin{table}[b]
\begin{tabular}{| l | c c c c c c c c|}
\hline
State & $g_0$ & $g_{+1}$ & $g_{+}2$ & $g_{+}3$ & $e_0$ & $e_{+1}$ & $e_{+2}$ & $e_{+3}$\\
Probability & 0.23 & 0.15 & 0.04 & 0.003 & 0.16 &0.09 & 0.02 &0.001\\
\hline
\end{tabular}
\caption{\label{tb:prob} Energy level occupation probabilities for one fixed atom. The parameters are: $g/\gamma=0.25$, $\kappa/\gamma=0.5$, $\mathcal{E}/\gamma=2$, and a Larmor frequency of $2.2\gamma/3$. The number of $V$-mode photons inside the cavity is $\langle a^\dagger_va_v\rangle\approx15$. Results for $g_{-1}$, $g_{-2}$, $g_{-3}$ and $e_{-1}$, $e_{-2}$, $e_{-3}$ follow by symmetry.}
\end{table}

\begin{figure}
\begin{center}
\includegraphics[width=0.90\linewidth]{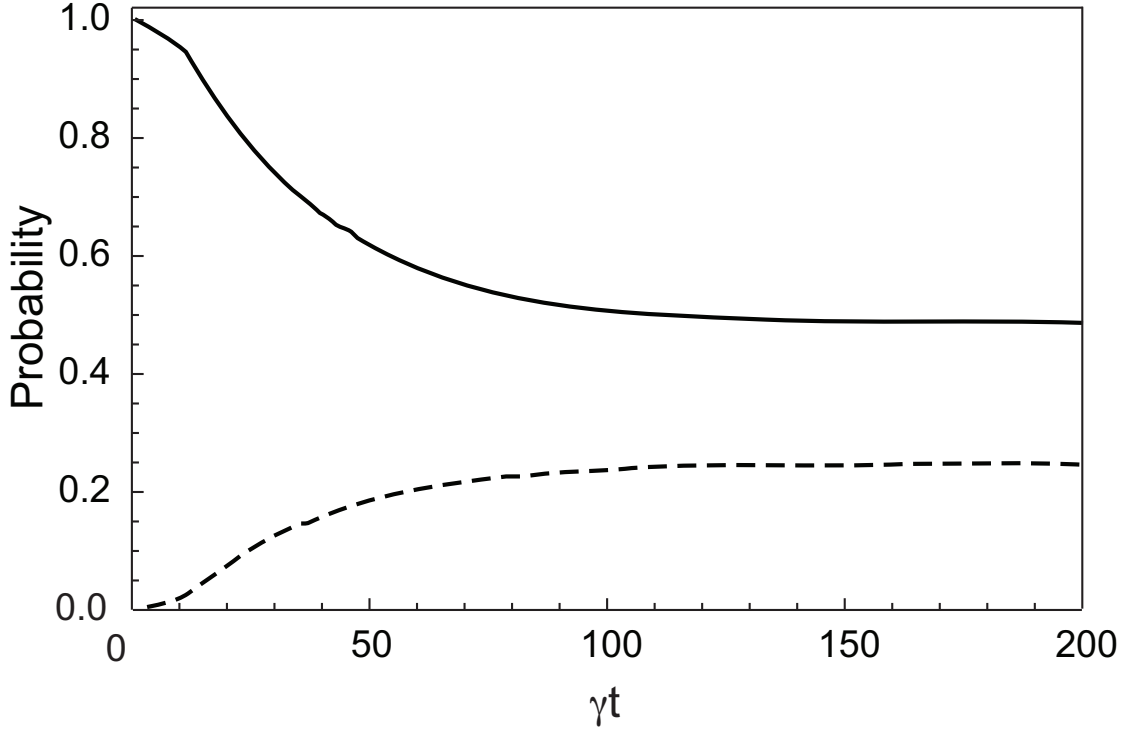}
\caption{\label{fig:prob_elements} Probability for the preparation of superposition state $|\psi_s\rangle$ (solid line) and either of the superposition states $|\psi_{\pm 1}\rangle$ (dashed line), for one fixed atom, as a function of the time of the first $H$-photon detection. The parameters are: $g/\gamma=0.25$,  $\kappa/\gamma=0.5$, $\mathcal{E}/\gamma=0.3$, and a Larmor frequency of $2.2\gamma/3$.}
\end{center}
\end{figure}

The distribution over atomic ground states prior to the detection of a first $H$-photon strongly effects the state the atom is projected into on average. If the time of the first detection is close to zero the atom is projected into $|\psi_0\rangle$ [Eq.~(\ref{eq:super0ground}]. As this time increases, superpositions of, first, $|g_0\rangle$ and $|g_{+2}\rangle$ (or $|g_{-2}\rangle$), and then  $|g_{\pm1}\rangle$ and $|g_{\pm3}\rangle$ appear. Numerically we have checked that the state immediately after the first $H$-photon  detection may be written approximately as
\begin{eqnarray}
\label{eq:afterstate}
\rho_c(t)&=&p_0|\psi_0\rangle\langle\psi_0|+p_1(|\psi_1\rangle\langle\psi_1|+|\psi_{-1}\rangle\langle\psi_{-1}|)\nonumber\\
&&+p_2 \Delta\rho_c,
\end{eqnarray}
with
\begin{eqnarray}
\label{eq:super1ground}
|\psi_{\pm 1}\rangle&=&(\sqrt{10}|g_{0}\rangle+\sqrt{3}|g_{\pm
  2}\rangle)/\sqrt{13}\, ,
\end{eqnarray}
where $p_i$, $i=0,1$, is the probability distribution over $|\psi_0\rangle$ and $|\psi_{\pm1}\rangle$, and $p_2=1-p_0-2p_1$ is the probability that the first $H$-photon is scattered out of one of the ground states with $|i|=2,3$; $\Delta\rho_c$ is the state reached from such scattering events. As can be seen from Fig.~\ref{fig:prob_elements}, as the wait for the first $H$-photon detection becomes longer, the probability to realize $|\psi_s\rangle$ decreases while that to realize $|\psi_{\pm1}\rangle$ increases. For the parameters considered, the sum $p_0+2p_1$ is close to 0.97, which tells us that $|\psi_s\rangle$ and $|\psi_{\pm1}\rangle$ cover all relevant prepared superpositions. An increase in drive strength changes only the time evolution of the probabilities, not their stationary values.

We move now to trajectories that deviate from the ideal after the first $H$-photon is detected [group (ii)]. We assume the prepared superposition is $|\psi_0\rangle$. Interaction of the atom with the driven $V$ cavity mode moves population to the superposition $|\psi_0^\prime(t)\rangle$ [Eq.~(\ref{eq:super0excited})]. It is then possible that spontaneous emission
(to the side) moves $|\psi_0^\prime(t)\rangle$ to a superposition of the ground states $|g_0\rangle$ and $|g_{+2}\rangle$ (or $|g_{-2}\rangle$). As the Clebsch-Gordan coefficient connecting $|e_{\mp1}\rangle$ to $|g_{0}\rangle$ differs from that
connecting $|e_{\pm1}\rangle$ to $|g_{\pm2}\rangle$), such an event yields unequal weights in the ground-state superposition. Nevertheless, this aside, the described process recovers the initial setup---a ground-state superposition---but in a manifold of states shifted to the right or left. Continuing then with the standard story, the $\pi$-polarized drive transfers the superposition of $|g_0\rangle$ and $|g_{\pm2}\rangle$ to a superposition of $|e_0\rangle$ and $|e_{\pm2}\rangle$ in the excited state. A second $H$-photon can then be emitted via the cavity, projecting the atom into a superposition of $|g_{\pm1}\rangle$, $|g_{\pm3}\rangle$ and $|g_{\mp1}\rangle$ [compare Eq.~(\ref{eq:return})]. Apart for the changed weight factors, the quantum eraser process still takes place, only within a different manifold of atomic states. Of course, the unequal weights yield a beat with diminished visibility.

Such spontaneous emission events can happen several times in the interval separating the $H$-photon detections; nevertheless, so long as there is no $\sigma_+$ ($\sigma_-$) emission at a time when $|e_{-3}\rangle$ ($|e_{+3}\rangle$) is part of the  superposition in the excited state, the superposition, with modified weights, survives. For an atom with six levels, as in Fig.~(\ref{eraser}), a single $\sigma_+$ or $\sigma_-$ spontaneous emission will destroy the prepared superposition, reducing the observed visibility far more than in the 16-level case. Our use of the 16-level configuration produces a particularly robust (against spontaneous emission) quantum beat.

\begin{figure}
\begin{center}
\includegraphics[width=0.90\linewidth]{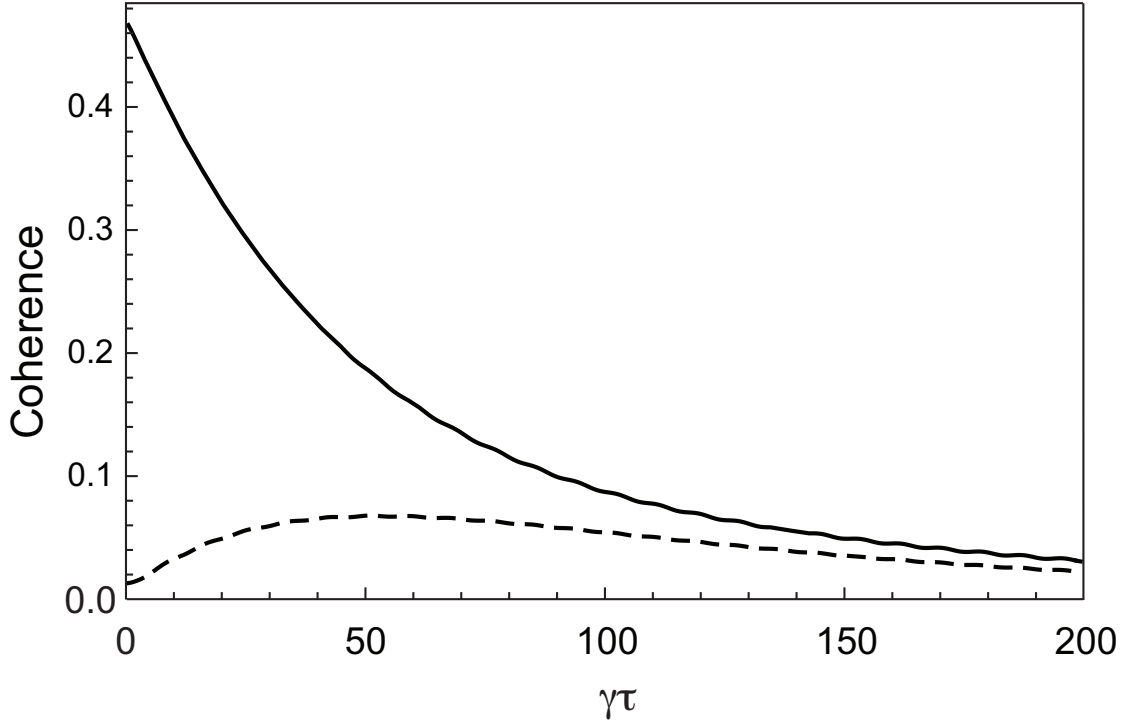}
\caption{\label{fig:prob_super}Evolution of spontaneously created coherences for one fixed atom; off-diagonal matrix elements $|\langle|g_{+1}|\rho(t+\tau) |g_{-1}\rangle|$ (solid line) and $|\langle g_{\pm2}|\rho(t+\tau)|g_0\rangle)|$ (dashed line) are plotted as a function of the time after a first $H$-photon is detected.  The parameters are: $g/\gamma=0.25$, $\kappa/\gamma=0.5$, $\mathcal{E}/\gamma=0.3$, and a Larmor frequency of $2.2\gamma/3$. The number of $V$-photons inside the cavity is approximately $0.3$.}
\end{center}
\end{figure}

Figure~\ref{fig:prob_super} shows how the spontaneously created coherence moves between different ground states after the first $H$-photon is detected. We quantify the coherence by off-diagonal matrix elements $|\langle g_{i+1}|\rho_c(t+\tau)|g_{i-1}\rangle|$, $i=0$ and 1, which fall between a maximum of 0.5 (equally weighted superposition) and zero. The solid line follows $|\langle g_{+1}|\rho(t+\tau)|g_{-1}\rangle|$ as a function of $\tau$. It begins at $\tau=0$ from  approximately $0.46$, which shows that immediately after the photon is detected the atom is to a good approximation in the superposition $|\psi_0\rangle$; $p_0\approx1$ in Eq.~(\ref{eq:afterstate}). As the time to the second photon detection progresses, $p_0$ decreases as $p_{\pm1}$ grows and part of the coherence is transferred to a superposition of $|g_{0}\rangle$ and $|g_{\pm 2}\rangle$. Eventually phase diffusion (decoherence) which accompanies repeated cycles of excitation and spontaneous emission sets in, causing both displayed coherences to decay to zero \cite{norris12}. Coherences between ground states other than those shown in the figure are negligible.

Before turning to the many atom case relevant to our experiment, one final effect might usefully be introduce at the one-atom level. The mirrors in the experiment show a small birefringence and mix a little of the $V$-polarized light with the $H$-polarized cavity mode. We attempt to null this mixing with a half-wave plate placed in the cavity output [Fig.~\ref{levels}(b)]; alternatively, in some measurements we deliberately enhance it (see Fig.~\ref{waveplate_mixing}). The mixing effectively performs a homodyne measurement with weak (at the one-photon level) local oscillator field. Let us make the substitution $a_h\to a_h+\epsilon$, where $\epsilon$ is the amplitude of the mixed drive light, taken for simplicity to be classical, real, and constant. The (unnormalized) intensity correlation function is now
\begin{eqnarray}
G^{(2)}(t,t+\tau)&=&G^{(2)}_h(t,t+\tau)\nonumber\\
&&+\epsilon^2\big\{G^{(1)}_h(t,t)+G^{(1)}_h(t+\tau,t+\tau)\nonumber\\
&&+2{\rm Re}[G_h^{(1)}(t,t+\tau)+G_h^{(a)}(t,t+\tau)]\big\}\nonumber\\
&&+\epsilon^4\, ,
\label{eq:gtwo_homodyne}
\end{eqnarray}
where $G^{(1)}_h(t,t+\tau)=\langle a_h^\dagger(t)a_h(t+\tau)\rangle$ is the first-order correlation function of the $H$ mode, and $G_h^{(a)}(t,t+\tau)=\langle a_h^\dagger(t)a_h^\dagger(t+\tau)\rangle$ is the $H$-mode anomalous correlation. Third-order correlations vanish for weak drive because scattering a second photon [blue wavy lines in Fig.~\ref{levels}(a)] leaves the atom in a manifold  orthogonal to that reached after scattering one photon [red wavy lines in Fig.~\ref{levels}(a)], i.e., one- and two-photon states entangle with orthogonal atomic states. The anomalous correlation survives because the manifold reached by scattering two photons contains the initial state---$|g_0\rangle$ in Fig.~\ref{levels}(a).

\begin{figure}[t]
\begin{center}
\includegraphics[width=0.90\linewidth]{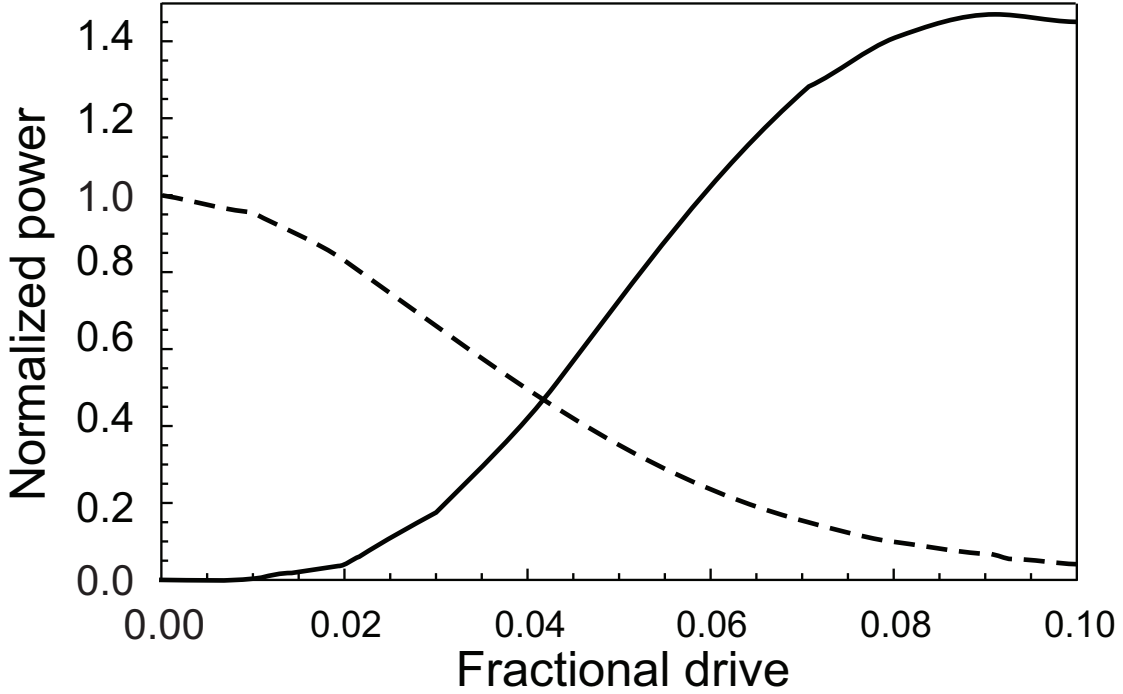}
\caption{\label{fig:ratiodu} Size of the beat at the Larmor frequency (solid line) relative to that at twice the Larmor frequency (dashed line), as a function of the fraction of drive light mixed with the $H$-mode. The parameters are: $g/\gamma=0.25$, $\kappa/\gamma=0.5$, $\mathcal{E}/\gamma=0.3$, and a Larmor frequency of $2.2\gamma/3$.
}
\end{center}
\end{figure}

With increasing $\epsilon$, the frequency of the quantum beat changes from twice the Larmor frequency to the Larmor frequency as the third term on the righ-hand side of Eq.~(\ref{eq:gtwo_homodyne}) comes to dominate the first. Figure~\ref{fig:ratiodu} illustrates the transition.

\subsection{Many atoms}\label{sc:manyatoms}
Our experiment is performed with a cold atomic beam and therefore the many-atom case must be considered. We generalize the non-Hermitian Hamiltonian of Eq.~(\ref{eq:hamiltonian}) by writing the free, interaction, and non-unitary loss terms for $N$ arbitrarily located atoms:
\begin{subequations}
\begin{eqnarray}
H_0&=&\hbar\omega_v a_v^\dagger a_v+\hbar\omega_h a_h^\dagger a_h\nonumber\\
\noalign{\vskip2pt}
&&+\sum_{j=1}^N\left(\sum_{i=-3}^3\hbar\omega_{g_i}|g_i\rangle^j\langle g_i|^j+\sum_{i=-4}^4\hbar\omega_{e_i}|e_i\rangle^j\langle e_i|^j\right),\nonumber\\
\noalign{\vskip-2pt}
&&{}\\
\noalign{\vskip-4pt}
H_I&=&\sum_{j=1}^N\hbar g_{j}[a_v^\dagger\Sigma_\pi^j+a_h^\dagger(\Sigma_{\sigma_+}^j+\Sigma_{\sigma_-}^j)]+{\rm h.c.}\, ,\\
H_{L}&=&-i\hbar\kappa(a_v^\dagger a_v+a_h^\dagger a_h)\nonumber\\
&&-\sum_{j=1}^Ni\hbar\frac{\gamma}{2}(\Sigma^{j\dagger}_\pi\Sigma^j_\pi+\Sigma^{j\dagger}_{\sigma_+}\Sigma^j_{\sigma_+}
+\Sigma^{j\dagger}_{\sigma_-}\Sigma^j_{\sigma_-}),
\end{eqnarray}
\end{subequations}
where the dipole coupling constants, $g_j$, $j=1,\ldots, N$, vary with the location of the atoms within the cavity mode function.

We aim to write the measured correlation function as a sum of terms applying to emission pathways for different atoms and their interference. To this end, we formally integrate the Heisenberg equation of motion for the $H$-mode annihilation operator, including its coupling to the reservoir. This yields (see, e.g., Ref.~\cite{carmichaelbookv2} page 206)
\begin{eqnarray}
\label{eq:cavity_dipole}
&a_h(t)=a_h(0)e^{-\kappa t}+\displaystyle{\sum_{j=1}^N}\mkern3mu g_j\displaystyle{\int_0^t}\Sigma^j_h(t-t')e^{-\kappa t^\prime}dt^\prime+{\rm v.f.},&\nonumber\\
\noalign{\vskip-8pt}
&&{}\\
\noalign{\vskip2pt}
&\Sigma^{i}_h(t)=\Sigma^{i}_{\sigma_-}(t)+\Sigma^{i}_{\sigma_+}(t),&
\end{eqnarray}
where v.f.\ indicates the presence of a vacuum field (reservoir) noise operator, which may be dropped from the calculation of a normal- and time-ordered average. We also drop the first term on the right-hand side of Eq.~(\ref{eq:cavity_dipole}), as we are interested in times much longer than the cavity decay time. Adding the mixed drive amplitude $\epsilon$, as above Eq.~(\ref{eq:gtwo_homodyne}), the $H$-mode cavity output is treated with the substitution
\begin{eqnarray}
\label{eq:cavity_dipole2}
&a_h(t)\to\displaystyle{\sum_{j=1}^N}\mkern2muA_j(t)+\epsilon,&\\
\noalign{\vskip4pt}
\label{eq:independent_source}
&A_j(t)=g_j\displaystyle{\int_0^t}\Sigma^j_h(t-t^\prime)e^{-\kappa t^\prime}dt^\prime.&
\end{eqnarray}
We further assume that (i) the probability that an emitted $H$-photon be re-absorbed before leaving the cavity is negligible and (ii) no additional $H$-photon is emitted in between the detection of photons at $t$ and $t+\tau$. The assumptions are justified, respectively, for moderate-to-weak dipole coupling and weak drive. They allow us to treat the atoms as independent and write the intensity correlation function as a generalization of Eq.~(\ref{eq:gtwo_homodyne}):
\begin{widetext}
\begin{eqnarray}
\label{eq:G2atoms_cavity}
G^{(2)}(t,t+\tau)&=&\sum_{j=1}^N\left(G^{(2)}_j(t,t+\tau)+\sum_{k\neq j=1}^NG^{(a)}_j(t,t+\tau)\Big(G^{(a)}_k(t,t+\tau)\Big)^{\mkern-2mu*}\right)\nonumber\\
&&+\sum_{j=1}^N\sum_{k\neq j=1}^N\left[G^{(1)}_j(t,t)G^{(1)}_k(t+\tau,t+\tau)+G^{(1)}_j(t,t+\tau)
\Big(G^{(1)}_k(t,t+\tau)\Big)^{\mkern-2mu*}\mkern2mu\right]\nonumber\\
&&+\epsilon^2\sum_{j=1}^N\mkern-2mu\Big\{G_j^{(1)}(t,t)+G_j^{(1)}(t+\tau,t+\tau)
+2{\rm Re}\Big[G_j^{(1)}(t,t+\tau)+G_j^{(a)}(t,t+\tau)\Big]\Big\}+\epsilon^4,
\end{eqnarray}
where we introduce individual atom correlation functions:
\begin{subequations}
\begin{eqnarray}
\label{eq:elements}
&G^{(2)}_j(t,t+\tau)=\langle A_j^\dagger(t)A_j^\dagger(t+\tau)A_j(t+\tau)A_j(t)\rangle,&\\
\noalign{\vskip2pt}
&G^{(1)}_j(t,t+\tau)=\langle A_j^\dagger(t)A_j(t+\tau)\rangle,\qquad
G^{(a)}_j(t,t+\tau)=\langle A_j^\dagger(t)A_j^\dagger(t+\tau)\rangle.&
\end{eqnarray}
\end{subequations}
\end{widetext}

In order to help with the interpretation of this expression, let us assume stationarity (correlation functions independent of $t$) and identical atoms. Equation (\ref{eq:G2atoms_cavity}) then reduces to
\begin{eqnarray}
\label{eq:gtwo_identical_atoms}
G^{(2)}(\tau)&=&NG^{(2)}_A(\tau)+N(N-1)|G^{(a)}_A(\tau)|^2\nonumber\\
&&+N(N-1)[I_A^2+|G^{(1)}_A(\tau)|^2]\nonumber\\
&&+2\epsilon^2N\{I_A+{\rm Re}[G^{(1)}_A(\tau)+G^{(a)}_A(\tau)]\}\nonumber\\
&&+\epsilon^4,
\end{eqnarray}
where $I_A=G^{(1)}_A(0)$ and the subscript denotes any atom. The result is similar to Eq.~(4) in \cite{carmichael78}. The only change is the anomalous correlation $G^{(a)}_A(\tau)$, which is retained here because we consider scattering into a single cavity mode, not free-space scattering as in \cite{carmichael78}; in the latter case, the random phases accompanying propagation from randomly located atoms to the detector cause the anomalous correlation to vanish. It is also important to recall that $G^{(a)}(\tau)$ is only nonzero because scattering two $H$-photons places the atom in a ground-state manifold that is not orthogonal to the initial state. $G^{(a)}(\tau)$ is therefore intimately related to the erasure of which-path information by the scattering of a second photon. Of course it approaches zero as $\tau\rightarrow\infty$, due to dephasing induced by spontaneous emission \cite{norris12}. The time scale for this is long compared with the transit time through the cavity if the drive is weak.

Leaving aside the homodyne terms, three correlation functions contribute to the quantum beat in Eq.~(\ref{eq:gtwo_identical_atoms}). There is first and foremost $G^{(2)}_A(\tau)$. It records the beat due to the interference of indistinguishable pathways for scattering two photons by one atom; there are $N$ such one-atom terms. There are then $N(N-1)$ copies of the first-order correlation function $|G^{(1)}_A(\tau)|^2$. These arise from the sum over cross-terms $G^{(1)}_j(\tau)[G^{(1)}_k(\tau)]^*$, $k\neq j$, in Eq.~(\ref{eq:G2atoms_cavity}). They also exhibit a beat at twice the Larmor frequency. It records the interference of indistinguishable pathways for scattering a first $H$-photon from atom $j$ ($k$) and a second from atom $k$ ($j$)---i.e., the interference of scattering events with reversed time-order, when both orders leave the same two atoms in the same final state. Finally, there are $N(N-1)$ copies of $|G_A^{(a)}(\tau)|^2$. These anomalous correlation functions record the many-atom extension of the $G^{(2)}_A(\tau)$ quantum beat. They are present because when one atom scatters two photons, our cavity setup is unable to tell from which of the $N$ atoms the two photons come.

\subsection{Atomic motion}
\label{sc:atomic_motion}

Our experiment is performed with a slow atomic beam. The atoms move through the cavity mode function and the dipole coupling coefficients, $g_j(t)$, $j=1,\ldots,N$, are randomly determined functions of time. If we neglect the effect of photon scattering on the atomic motion, $g_j(t)$, for a particular atom, is defined by a (constant) velocity ${\bm v}_j$, a time, $t_j$, at which the atom crosses the plane perpendicular to the beam containing the cavity axis, and a position ${\bm r}_j$ on that plane; the velocity may be further specified by a speed $v_j$, and polar and azimuthal angles, $\theta_j$ and $\phi_j$, defined with respect to the cavity axis. The quantity relevant for our experiment is the time average
\begin{equation}
G^{(2)}(\tau)=\frac1T\int_0^TG^{(2)}(t,t+\tau) dt\,
\end{equation}
of Eq.~(\ref{eq:G2atoms_cavity}); we need also, for normalization, the mean intensity
\begin{equation}
\label{eq:time_averaged_intensity}
I=\frac1T\int_0^T\sum_{j=1}^NG^{(1)}_j(t,t)dt+\epsilon^2.
\end{equation}
The sums in Eq.~(\ref{eq:G2atoms_cavity}) may be considered to range over every atom that enters the cavity during the course of the experiment. The atoms are uniformly distributed in a beam of rectangular cross section, height $d$ and extension $l$ along the cavity axis, at flux density $F$ (number of atoms  per second through unit area). Correlation functions for atom $j$ depend on ${\bm r}_j$, ${\bm v}_j$, and the time difference $t-t_j$. They fall to zero for $|t-t_j|$ much larger than the mean transit time across the cavity, $\tau_0=w_0/\langle v\rangle$, where $w_0$ is the mode function waist and $\langle v\rangle$ is the mean speed of an atom along the axis of the atomic beam.

We consider the mean intensity as an illustration of the way to proceed and then pass directly to the result for $G^{(2)}(\tau)$. For the time integral on the right-hand side of Eq.~(\ref{eq:time_averaged_intensity}), we may write (with $t^\prime=t-t_j$)
\begin{equation}
\int_0^T\sum_{j=1}^NG^{(1)}_j(t,t)dt=FT\ell d\left\langle\int_{-\infty}^\infty G^{(1)}_{{\bm r}_j,{\bm v}_j,t_j}(t^\prime,t^\prime)dt^\prime\right\rangle,
\end{equation}
where it is clear that the number of nonzero contributions arising from the sum is just the mean number of atoms, $FT\ell d$, crossing the plane containing the cavity axis during time $T$, while the angle bracket denotes an ensemble average, over ${\bm r}_j$ and ${\bm v}_j$, for atoms distributed within the beam cross-section; the ensemble average can be taken numerically. We then define a mean intensity per (effective) atom normalized to the profile of the mode function,
\begin{equation}
I_A=\frac{\ell d}{\pi\ell w_0/4}\frac1{\tau_0}\left\langle\int_{-\infty}^\infty G^{(1)}_{{\bm r}_j,{\bm v}_j,t_j}(t^\prime,t^\prime)dt^\prime\right\rangle,
\end{equation}
and arrive at
\begin{equation}
\label{eq:time_averaged_intensity_again}
I=\bar N_{\rm eff}I_A+\epsilon^2,
\end{equation}
where $\bar N_{\rm eff}=\rho\pi w_0^2\ell/4$ is the effective number of atoms (see Ref.~\cite{carmichael99}, for example), with $\rho=F/\langle v\rangle$ the atomic density.

The time average of each sum in Eq.~(\ref{eq:G2atoms_cavity}) is treated in a similar way. This yields a straightforward generalization of Eq.~(\ref{eq:gtwo_identical_atoms}):
\begin{eqnarray}
\label{eq:quasefinal}
G^{(2)}(\tau)&=&\bar N_{\rm eff}G^{(2)}_A(\tau)+\bar N_{\rm eff}^2|G^{(a)}_A(\tau)|^2\nonumber\\
&&+\bar N_{\rm eff}^2[I_A^2+|G^{(1)}_A(\tau)|^2]\nonumber\\
&&+2\epsilon^2\bar N_{\rm eff}\{I_A+{\rm Re}[G^{(1)}_A(\tau)+G^{(a)}_A(\tau)]\}\nonumber\\
&&+\epsilon^4,
\end{eqnarray}
with ($\xi=1,2,a$)
\begin{equation}
\label{eq:time_averaged_correlation_functions}
G_A^{(\xi)}(\tau)=\frac{\ell d}{\pi\ell w_0/4}\frac1{\tau_0}\left\langle\int_{-\infty}^\infty G^{(\xi)}_{{\bm r}_j,{\bm v}_j,t_j}(t^\prime,t^\prime+\tau)dt^\prime\right\rangle.
\end{equation}
We compare experimental results with a numerical evaluation of Eq.~(\ref{eq:quasefinal}) in Section~\ref{sc:exp-theo}. In the next subsection we introduce simplifications that lead to a closed expression for $G^{(2)}(\tau)$.

\subsection{Bad-cavity and adiabatic limit}
We make two simplifying assumptions and focus on the weak-field limit. First, we assume that the cavity decay rate $\kappa$ is sufficiently large compared with all other rates that the dipole operator may be taken outside the integral in Eq.~(\ref{eq:independent_source}) (bad-cavity limit \cite{carmichaelbookv2}), allowing us to write $A_j(t)=g_j(t)\Sigma^j_h(t)/\kappa$. We assume also that the atomic motion is slow and atomic states follow the changing coupling constants, $g_j(t)$, $j=1,\ldots,N$, adiabatically. Then noting that the dipole operator is expected to carry an atomic excitation proportional to $g_j(t)$ in the weak-field limit, which we verify numerically, we may factor out the $g_j(t)$-dependence and write each individual atom correlation function in the form
\begin{equation}
\label{eq:adiabatic_correlation_functions}
G^{(\xi)}_{{\bm r}_j,{\bm v}_j,t_j}(t,t+\tau)=\left[\frac{g_{\rm max}}{\kappa}\bar g_j(t)\bar g_j(t+\tau)\right]^{i_\xi} G^{(\xi)}_{\rm max}(\tau),
\end{equation}
with $i_\xi=2,4,2$ for $\xi=1,2,a$, where $G^{(\xi)}_{\rm max}(\tau)$ is the correlation function, in the long-time limit, for a fixed atom at maximum coupling strength $g_{\rm max}$. The scaled coupling functions, $\bar g_j(t)$, $j=1,\ldots,N$, follow from the cavity mode function:
\begin{equation}
\bar g_j(t)=\cos[kz_i(t)]e^{-[x_j(t)^2+y_j(t)^2]/w_0^2}\, ,
\end{equation}
where $k=2\pi/\lambda$ is the wavenumber, and $x_j(t)$, $y_j(t)$, $z_j(t)$ define the trajectory of atom $j$. We assign the $z$-axis parallel to the cavity axis and the $x$-axis as the axis of the atomic beam; at time $t^\prime=t-t_j=0$, we identify ${\bm r}_j=(y_j(t_j),z_j(t_j))$, and ${\bm v}_j=(\dot x_j(t),\dot y_j(t),\dot z_j(t))$.

With these simplifications, the time integration and average of Eq.~(\ref{eq:time_averaged_correlation_functions}) are defined entirely by the prescribed coupling functions $\bar g_j(t)$, $j=1,\ldots,N$, and the velocity distribution of the atomic beam. In some simple cases analytical results can be obtained. The first is when the atoms move along parallel trajectories perpendicular to the cavity and at a common speed $v$. Equations (\ref{eq:time_averaged_correlation_functions}) and (\ref{eq:adiabatic_correlation_functions}) then yield
\begin{subequations}
\begin{equation}
G^{(2)}_A(\tau)=\frac{35}{256}\frac{g_{\rm max}^4}{\kappa^4}e^{-2\tau^2/\tau_0^2}G_{\rm max}^{(2)}(\tau),
\end{equation}
and
\begin{equation}
G_A^{(1,a)}(\tau)=\frac38\frac{g_{\rm max}^2}{\kappa^2}e^{-\tau^2/\tau_0^2}G^{(1,a)}_{\rm max}(\tau),
\end{equation}
\end{subequations}
and setting $\epsilon=0$, for simplicity, and introducing the normalized correlation functions
\begin{subequations}
\begin{equation}
g^{(2)}(\tau)=\frac{G^{(2)}(\tau)}{(\bar N_{\rm eff}I_A)^2},\qquad g^{(2)}_{\rm max}(\tau)=\frac{G^{(2)}_{\rm max}(\tau)}{[G^{(1)}_{\rm max}(0)]^2},
\end{equation}
and ($\xi=1,a$)
\begin{equation}
g^{(\xi)}_{\rm max}(\tau)=\frac{G^{(\xi)}_{\rm max}(\tau)}{G^{(1)}_{\rm max}(0)},
\end{equation}
\end{subequations}
we arrive at a correlation function with Gaussian transit time decay:
\begin{eqnarray}
\label{eq:fixed_speed_correlation_function}
g^{(2)}(\tau)&=&1+e^{-2\tau^2/\tau_0^2}\left[\vphantom{\frac{35}{36}}|g^{(1)}_{\rm max}(\tau)|^2+|g^{(a)}_{\rm max}(\tau)|^2\right.\nonumber\\
&&\left.+\frac{35}{36}\bar N_{\rm eff}^{-1}g^{(2)}_{\rm max}(\tau)\right].
\end{eqnarray}
Note how the dominant term is $g^{(2)}_{\rm max}(\tau)$ (interfering pathways for two photons emitted by one atom) at small values of $\bar N_{\rm eff}$, less than one, whereas for $\bar N_{\rm eff}\gg 1$, $|g^{(1)}_{\rm max}(\tau)|^2$ (interfering time orders for two photons emitted by different atoms) and $|g^{(a)}_{\rm max}(\tau)|^2$ (interfering pathways for two photons emitted by one atom or by another) are the dominant terms.

\begin{figure}[t]
\begin{center}
\includegraphics[width=0.90\linewidth]{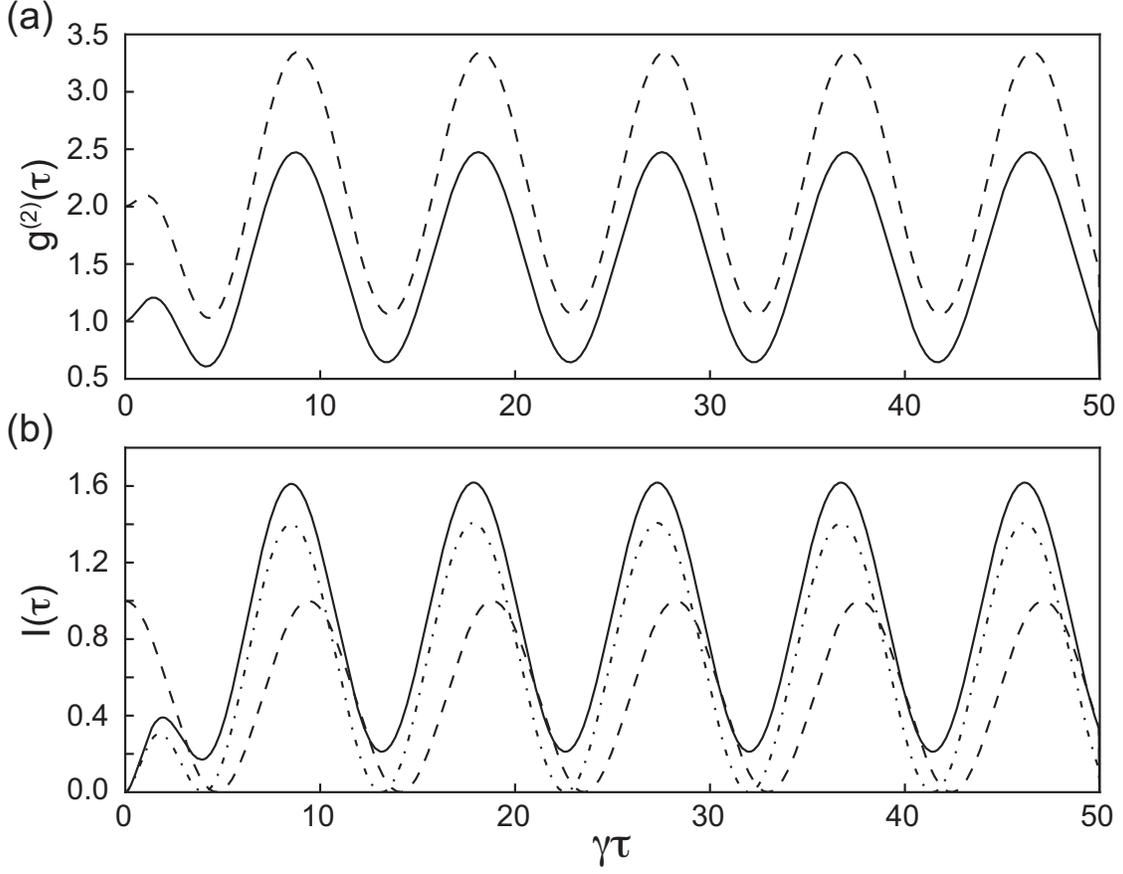}
\caption{\label{fig:descomposition16} (a) Quantum beat plotted from Eq.~(\ref{eq:fixed_speed_correlation_function}) in the fixed atom limit ($\tau_0\to\infty$): $\bar N_{\rm eff}=35/36$ (solid) and $\bar N_{\rm eff}\gg1$ (dashed). (b) Intensity $I(\tau)$ of the three terms that contribute to make up the oscillatory part of the solid curve in (a): $g^{(2)}_{\rm max}(\tau)$ (solid), $|g^{(1)}_{\rm max}(\tau)|^2$ (dashed), and $|g^{(a)}_{\rm max}(\tau)|^2$ (dot-dashed). The parameters are: $g_{\rm max}/\gamma=0.005$, $\kappa/\gamma=0.5$,  $\mathcal{E}/\gamma=0.005$, and a Larmor frequency of $\gamma/3$}.
\end{center}
\end{figure}

To illustrate the quantum beat we recover the case of fixed atoms (averaged over locations) by taking $\tau_0\to\infty$. Two examples are shown in Fig.~\ref{fig:descomposition16}(a): the first with all three terms making equal contributions to the beat, and the second where $|g^{(1)}_{\rm max}(\tau)|^2$ and $|g^{(a)}_{\rm max}(\tau)|^2$ dominate.  Figure \ref{fig:descomposition16}(b) illustrates the behavior of each term separately for the former case. All terms oscillate with the same frequency---twice the ground-state Larmor frequency---in the weak-drive limit. There is a phase difference between the oscillations of $g^{(2)}_{\rm max}(\tau)$ and $|g^{(a)}_{\rm max}(\tau)|^2$ (two-photon amplitudes) and those of $|g^{(1)}_{\rm max}(\tau)|^2$ (one-photon amplitudes). This arises from different gyromagnetic ratios in the ground and excited states.

A more realistic modeling of our experiment takes the speed distribution, $D(v)$, of the atoms to correspond to a thermal effusive source: $D(v)=2\alpha^{-4}v^3 e^{-v^2/\alpha^2}$, where $\alpha^2=2k_BT/m$ \cite{ramsey1985molecular}. The additional average over speed yields
\begin{eqnarray}
\label{eq:variable_speed_correlation_function}
g^{(2)}(\tau)&=&1+\left(1+\frac\pi4\frac{\tau^2}{\tau_0^2}\right)^{\mkern-3mu-3}\mkern-3mu\left(|g^{(1)}_{\rm max}(\tau)|^2+|g^{(a)}_{\rm max}(\tau)|^2\right)\nonumber\\
&&+\frac{35}{36}\bar N_{\rm eff}^{-1}\mkern-3mu\left(1+\frac\pi2\frac{\tau^2}{\tau_0^2}\right)^{\mkern-3mu-3/2}\mkern-3mug^{(2)}_{\rm max}(\tau),
\end{eqnarray}
where $\tau_0=w_0/\langle v\rangle$ is defined with $\langle v\rangle=\sqrt{8k_BT/\pi m}$---the mean speed of an atom in the source.

\begin{figure}[h!]
\includegraphics[width=0.90\linewidth]{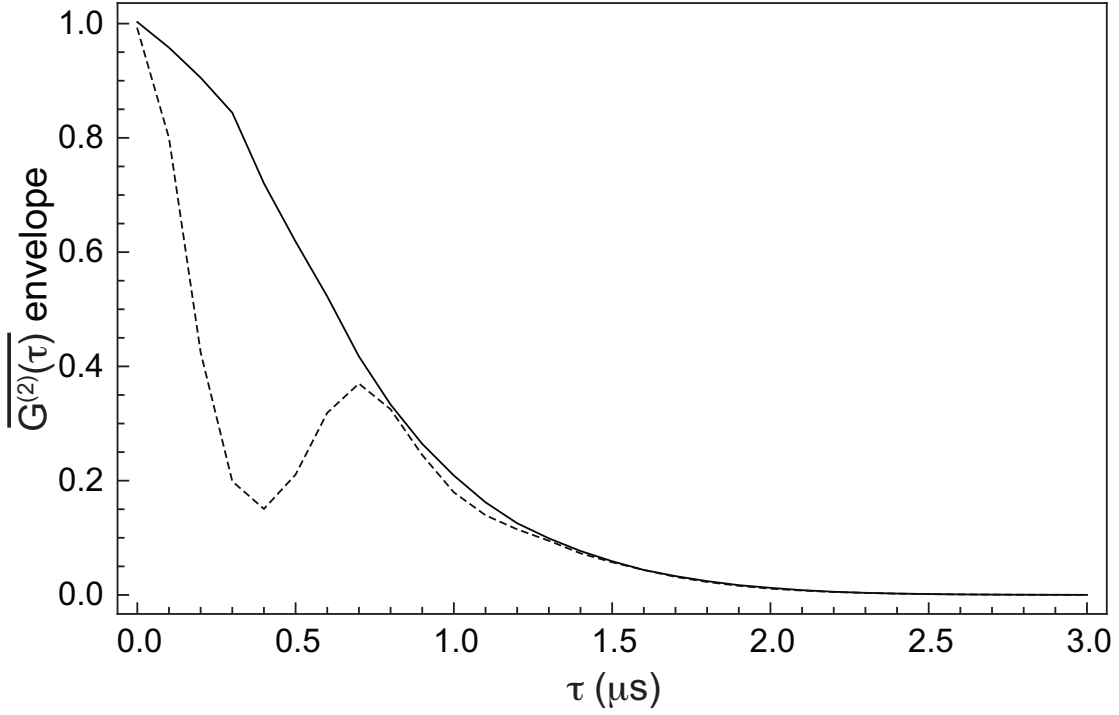}
\caption{\label{fig:envelope} Effect of a mean tilt of the atomic beam relative to the cavity axis. The envelope of $g^{(2)}(\tau)-1$ is plotted for no tilt of the beam (solid) and a mean tilt  $\langle\theta\rangle=1.3\mkern2mu{\rm deg}$ (dashed), with $\langle v\rangle=22\mkern2mu{\rm m/s}$, $\Delta v=2\mkern2mu{\rm m/s}$, and $\Delta\theta=0.025\mkern3mu{\rm deg}$.}
\end{figure}

In practice the atoms do not move perfectly perpendicular to the cavity axis. We do not have an analytical expressions for this most general case. Figure~\ref{fig:envelope} shows the numerically calculated envelope of the correlation function for the case $\bar N_{\rm eff}\gg1$, assuming a triangular distribution for the polar angle relative to the cavity axis $\theta$ and a Maxwell-Boltzmann speed distribution.  The figure shows how the standing-wave structure of the cavity mode function becomes imprinted on the envelope of the quantum beats. The local minimum of the dotted curve corresponds to the delay time when a majority of atoms pass from an anti-node to a node between the detection of the first and second photons. The spread in angle and speed explains why this structure does not recur at longer delays.

\section{Experiment and Results}
\label{sc:experiment}

\subsection{Apparatus}
We perform measurements using a slightly modified version of the apparatus described in Ref.~\cite{norris09a}. A sketch of the experimental setup appears in Fig.~\ref{levels}(b).  We probe a small ensemble of $^{85}$Rb atoms coupled to a Fabry-Perot resonator in vacuum.  The $2.2\mkern2mu{\rm mm}$ cavity has a $56\mkern2mu\mu{\rm m}$ mode waist and a finesse of $11,000$, with losses shared approximately equally between mirror transmission and scattering/absorption.  The decay rates of the field and atomic dipole---$(\kappa,\gamma/2)/2\pi=(2.8,3.0)\times10^6\mkern2mu{\rm s}^{-1}$---are approximately matched, and twice as large as the dipole coupling strength, $g_{\rm max}/2\pi=1.5\mkern2mu{\rm MHz}$, on the $D_2$-line $F=3$, $m=0$ to $F^\prime=4$, $m^\prime=0$  transition; this places our system in the intermediate coupling regime of cavity QED---single-atom cooperativity $C_1=g^2/\gamma\kappa=0.12$ and saturation photon number $n_0=\gamma^{2}/3g^2=5.3$---with only a small probability of reabsorption after a photon is emitted into the cavity mode.

A crossed-polarizer configuration separates the weak $H$-mode fluorescence from the much stronger $V$-polarized drive, necessitating careful selection and alignment of polarization elements.  We drive the cavity with a laser sideband generated by a polarization-maintaining single-mode $780\mkern2mu{\rm nm}$ fiber modulator (EO Space) operated at $230\mkern2mu{\rm MHz}$.  Before entering the mode-matching lens and vacuum chamber, the drive is linearly polarized---extinction ratio less than $5\times10^{-5}$---after passing through a Glan-Thompson polarizer and zero-order half-wave plate (HWP).  A second zero-order HWP placed after the cavity aligns the polarization to a calcite Wollaston prism for separation of the $H$- and $V$-mode light at the output.  The extinction ratio after this splitter is limited by birefringence in the cavity mirrors, vacuum chamber windows and lenses.  Its exact value is a function of drive intensity, likely due to thermoelastic stress-induced birefringence in the components, but is generally of order $5\times10^{-4}$.  The splitting of $H$- and $V$-mode resonance frequencies due to cavity mirror birefringence is less than $200\mkern2mu{\rm kHz}$.

The separated beams go to two avalanche photodiodes (APDs, Perkin-Elmer) for photon counting, except when measuring autocorrelation functions, in which case the $V$-mode beam is blocked and the $H$-mode beam is split equally between the two detectors by means of a separate HWP and polarizing beam splitter.  The TTL output channel of each detector is electronically split between a counter unit, for measuring rates, and a PC time-stamp card (Becker and Hickl DPC-230) for recording detection times with $164\mkern2mu{\rm ps}$ resolution.  A typical time series measurement takes approximately $300\mkern2mu{\rm s}$. The output of the time-stamp card is written to a text file and parsed by a C++ program to calculate the cross-correlation between detection events (the autocorrelation of the $H$-mode).  The correlated events are recorded in a histogram, bin width 16.4 ns, extending out to $\pm$16.4 $\mu$s.  Division by the average (uncorrelated) bin count yields $g^{(2)}(\tau)$.  The power spectrum is calculated from a discrete fast fourier transform (FFT) of this function.

\begin{figure}
\begin{center}
\includegraphics[width=0.80\linewidth]{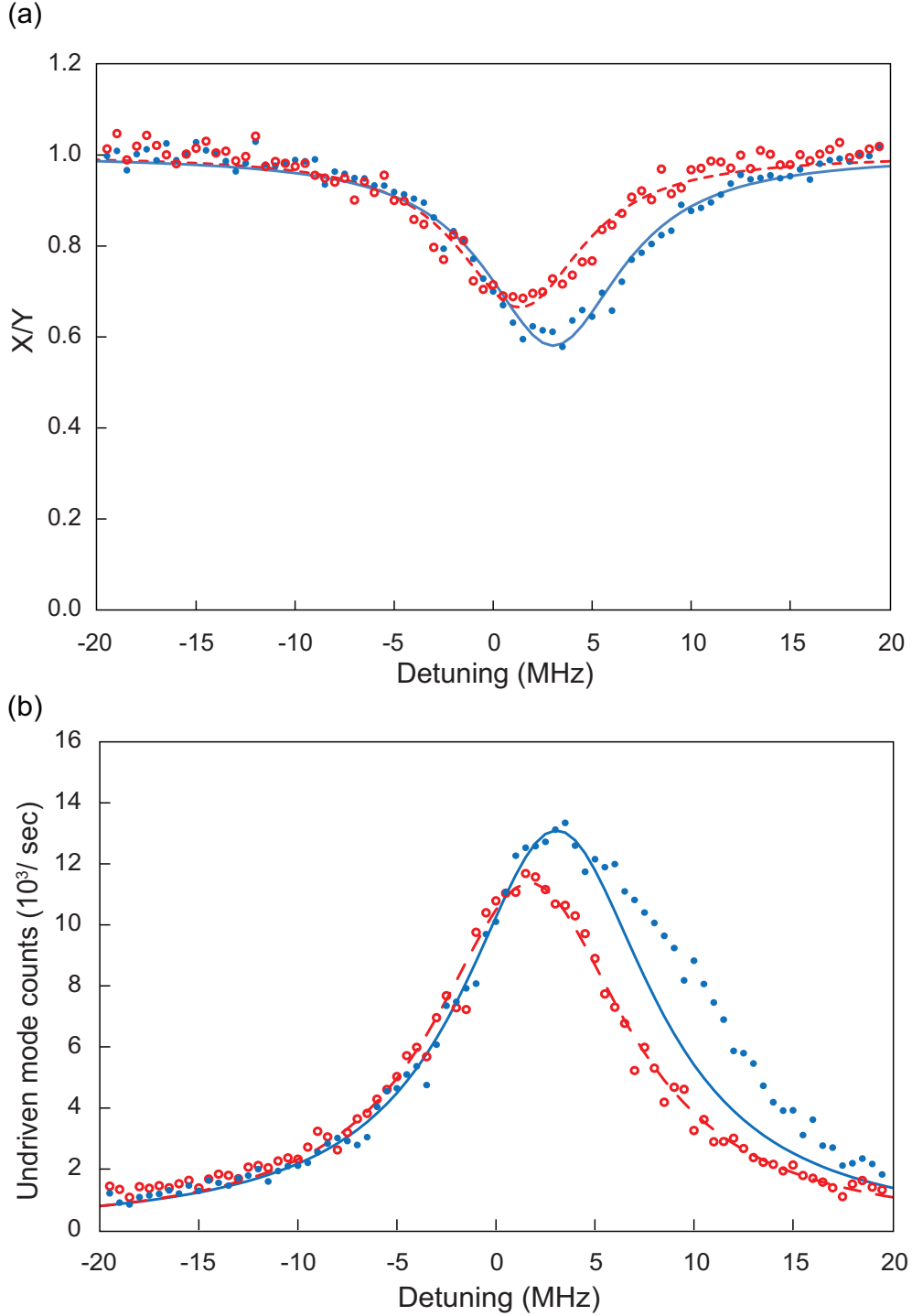}
\caption{(Color online) (a) Example of the $V$-mode absorption fraction as a function of common laser and cavity detuning from the central $m=0$ to $m^\prime=0$ atomic transition; for a magnetic field  of $\sim5\mkern2mu{\rm G}$ and approximately one photon on average in the cavity. The optical pumping beam (red/open circles ``on'', blue/filled dots ``off'') shifts the population toward the center. Solid lines are least-square fits of the data to a Lorentzian line shape. (b) $H$-mode count rates measured concurrently with the data in (a).
\label{opticalpumping}}
\end{center}
\end{figure}

The $^{85}{\rm Rb}$ atoms are extracted continuously from a magneto-optical trap (MOT) operating as a low-velocity intense source (LVIS) \cite{lu96} in a chamber directly above the cavity.  The atoms have a mean speed of $\sim22\mkern2mu{\rm m/s}$, which yields an interaction time of a few microseconds.  Atoms leaving the MOT are  primarily pumped to the $m=3$ ground state, with quantization axis provided by the residual vertical magnetic field from the MOT ``anti-Helmholtz''coil pair ($\sim7\mkern2mu{\rm G}$ at the location of the cavity). In order to change the magnetic field in the cavity, a third coil is added directly below the ``anti-Helmholtz" pair, with separate currents applied to each of the three coils. We are thus able to vary the vertical magnetic field in the cavity between $\pm12\mkern2mu{\rm G}$ while maintaining the required gradient for the MOT.  An additional pair of coils is oriented with axis parallel to the cavity axis to cancel any residual field in that direction.

Before atoms enter the cavity mode they are optically pumped to the $m=0$ ground state using a beam resonant with the $F=3$ to $F^\prime=3$ transition and polarization parallel to the vertical magnetic field.  The optical pumping beam is combined with light from the MOT re-pumper laser in a 50/50 polarization-maintaining single-mode fiber splitter, and collimated to a waist diameter of $0.5\mkern2mu{\rm mm}$ in order to pass between the top of the cavity mirrors and the upper edge of the vacuum window.  Due to strong scattering from multiple reflections into the APDs, we are unable to use the beam in a retro-reflected configuration.  It therefore imparts a net momentum kick to the atoms.  The intensity of the optical pumping beam is chosen optimally as a compromise between moving most atoms to $m=0$, while not ejecting too many from the beam, and scattering too strongly
into the APDs.

Figure \ref{opticalpumping} shows a typical measurement sequence used for optimizing the optical pumping configuration.  Frame (a) shows the absorption fraction (output intensity over input intensity, denoted $X/Y$) as measured from the $V$-mode count rates with the cavity and drive frequencies simultaneously swept across the atomic resonance.  The effect of the optical pumping is to shift and narrow the absorption peak, ideally to yield a symmetric lineshape centered around the resonance frequency of the $F=3$, $m=0$ to $F^\prime=4$, $m^\prime=0$ transition ($0\mkern2mu{\rm MHz}$ in the plot).  The departure of the center frequency from zero results from a combination of incomplete optical pumping and a small drift
in the frequency setpoint of the Rb saturated absorption spectroscopy reference used for the laser.  In frame (b) count rates for the $H$-mode light show a similar effect.  We use the center frequency obtained from these scans as the reference (zero-detuning) point for
our measurements.

The extensive Optical Bistability literature is useful for understanding and interpreting the effects of absorption and detuning in our experiment. When making these connections (see for example \cite{lugiato84}) it should be noted that we operate in the low intensity limit, and simultaneously scan the laser and cavity in order to address the atoms directly.

\subsection{Results}
\label{sc:results}

\begin{figure}
\begin{center}
\includegraphics[width=0.90\linewidth]{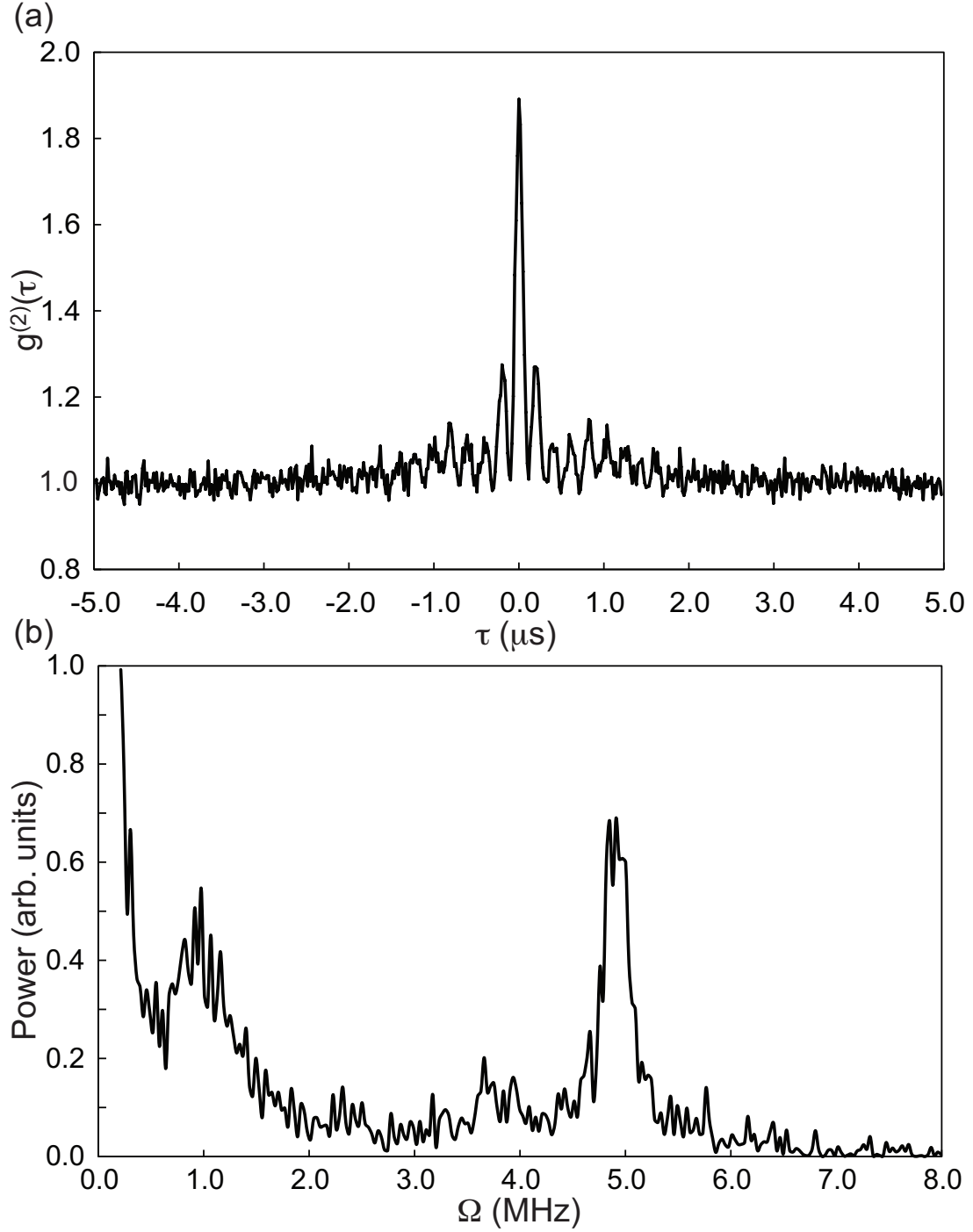}
\caption{(a) Measured intensity correlation function $g^{(2)}(\tau)$ and (b) its FFT power spectrum; for a $5\mkern2mu{\rm G}$ magnetic
field, $\bar N_{\rm eff}=2.9$, and approximately $6.5$ photons in the $V$ mode with no atoms present. The peak in the spectrum is located at $\approx 4.8\mkern2mu{\rm MHz}$, twice the ground state Larmor frequency for ${}^{85}{\rm Rb}$ in a $5\mkern2mu{\rm G}$ field.}
\label{beatsandFT}
\end{center}
\end{figure}

Frame (a) of Fig.~\ref{beatsandFT} displays a measured correlation function for a magnetic field of $5\mkern2mu{\rm G}$; its calculated power spectrum is displayed in frame (b). The main peak near 4.8 MHz corresponds to the quantum beat resonance. A smaller peak at half this frequency is also present, though largely obscured by noise.  It is the result of homodyne interference with drive light mixed in by cavity birefringence [see the paragraph surrounding Eq.~(\ref{eq:gtwo_homodyne})].  The small sidebands on the main peak correspond to a slight modulation of the beat envelope. The modulation is visible in frame (a) and results from the small ($1-2$ degree) deviation of the atomic beam from normal incidence with the axis of the cavity, which introduces sinusoidally varying coupling strengths, $g_j(t)$, and amplitude modulation of the spontaneous emission rate (see Fig.~\ref{fig:envelope}).

\begin{figure}
\begin{center}
\includegraphics[width=0.90\linewidth]{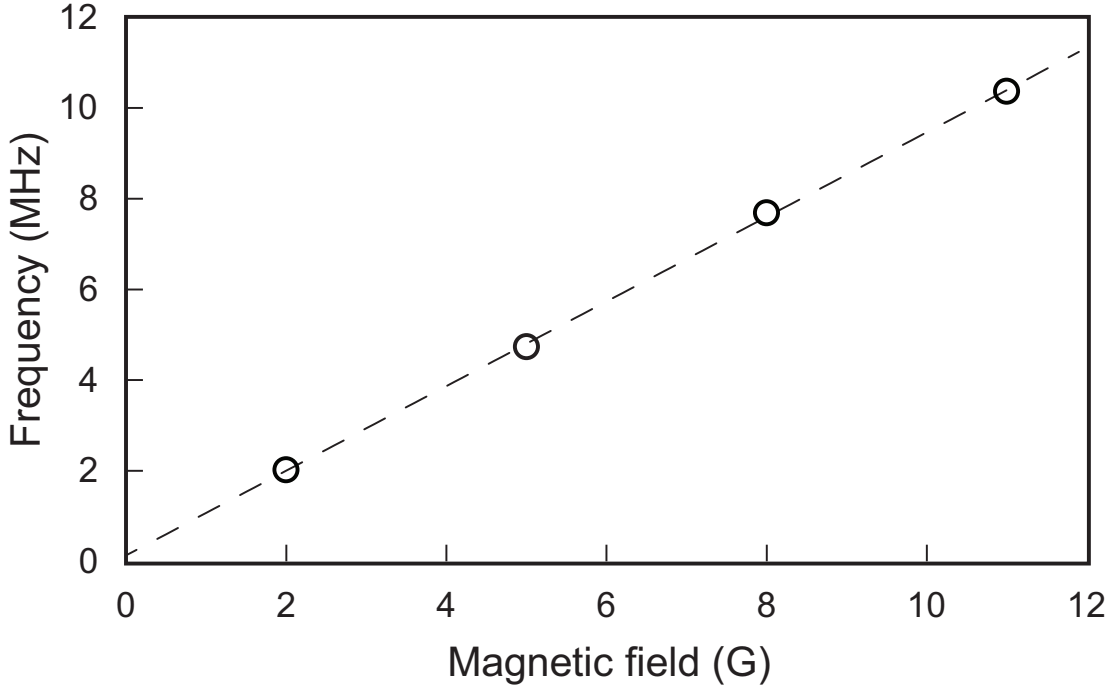}
\caption{Measured linear dependence of the frequency of the quantum beat on magnetic field.\label{frequencyvsmagfield}}
\end{center}
\end{figure}

Figure \ref{frequencyvsmagfield} illustrates the changing frequency of the quantum beat with increasing magnetic field, where the expected linear dependence is observed.  We note that the beat frequencies also depend on the intensity of the drive through an anomalous light-shift, which we report elsewhere \cite{norris12}.  Those presented here are extracted as the zero-intensity (i.e.~unshifted) limit of the measured frequencies for each magnetic field.  Our zeroing of the magnetic field in all three directions is not better than $10\mkern2mu{\rm mG}$. This error in independent calibration makes the small offset in the figure consistent with zero.

\begin{figure}
\begin{center}
\includegraphics[width=0.80\linewidth]{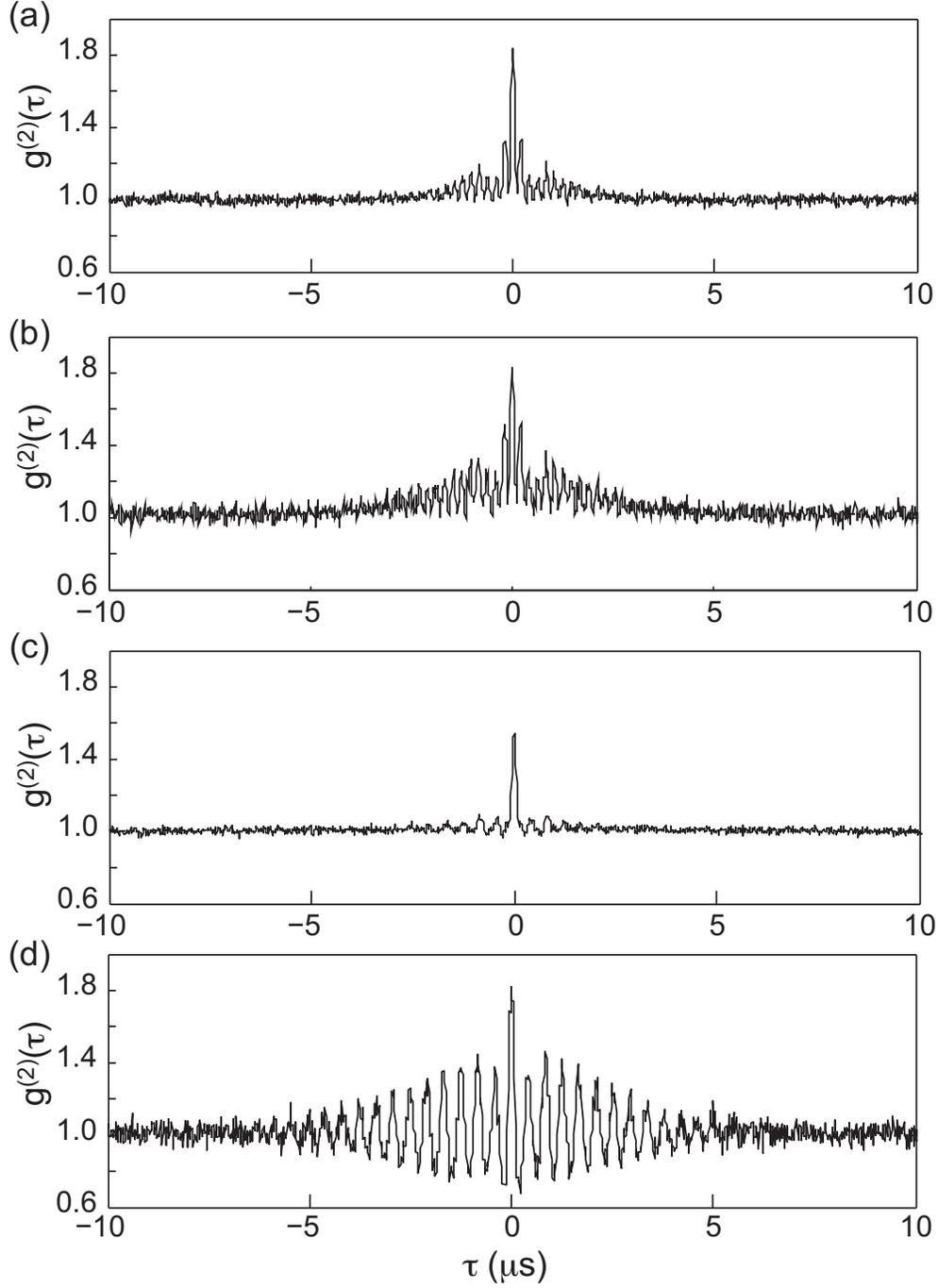}
\caption{Evolution of the measured $g^{(2)}(\tau)$ with homodyne mixing and detuning of the drive:  (a) on resonance, maximum extinction; (b) $-6\mkern2mu{\rm MHz}$ detuning, maximum extinction; (c) on resonance, HWP rotated by $2.8$ degrees; (d) $-6\mkern2mu{\rm MHz}$ detuning, HWP rotated by $1.2$ degrees. Data taken for a $5\mkern2mu{\rm G}$ magnetic field, $\bar N_{\rm eff}=2.9$, and approximately $2.0$ photons in the $V$ mode with no atoms present. \label{waveplate_mixing}}
\end{center}
\end{figure}

Figure \ref{waveplate_mixing} illustrates the change in the observed beat when the polarization presented to the detector is not taken orthogonal to that of the drive but allowed to rotate by a few degrees.  The rotation is controlled by changing the angle of the HWP placed between the cavity and the PBS (Fig.~\ref{levels}b). This mixes a small amount of drive light with the scattered light. With increasing mixed-in fraction the beat is eventually dominated by a homodyne term, which arises from the correlation of a photon scattered into the $H$ mode with a photon from the drive [see Fig.~\ref{fig:ratiodu} and terms proportional to $\epsilon ^2$ in Eqs.~(\ref{eq:gtwo_homodyne}), (\ref{eq:G2atoms_cavity}), (\ref{eq:gtwo_identical_atoms}), and (\ref{eq:quasefinal})]; thus, as in the two-atom case, interfering time orders also yield a quantum beat. This beat oscillates at half the frequency and allows the correlation function to dip below one.  Generally, some drive light is coupled into the $H$ mode through a small birefringence of the cavity mirrors.  Frames (c) and (d) of Fig.~\ref{waveplate_mixing} are recorded at the HWP angle that gives maximum visibility of the half-frequency beat.  Frames (b) and (d) show the dramatic increase in visibility gained when the driving laser is slightly detuned from resonance.  This is related to a decreased decoherence rate from quantum jumps (see Ref.~\cite{norris12}).

\begin{figure}
\begin{center}
\includegraphics[width=0.90\linewidth]{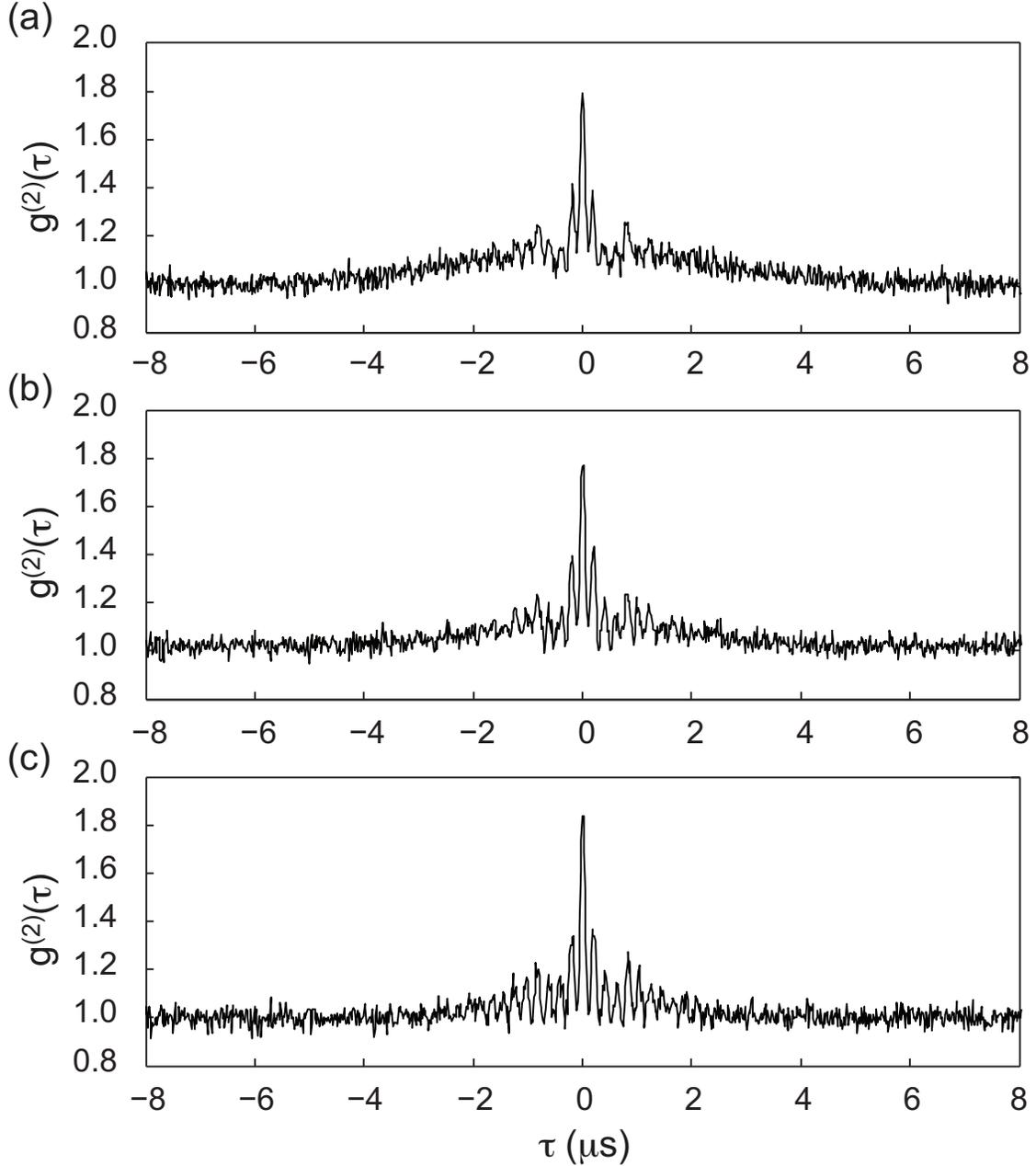}
\caption{Evolution of the measured $g^{(2)}(\tau)$ with increasing atomic beam density. Data taken for a $5\mkern2mu{\rm G}$ magnetic field, $\bar N_{\rm eff}=0.3, 0,9, 2.9$, (a,b,c) and approximately $3.1$ photons in the $V$ mode with no atoms present.\label{atomnumber}}
\end{center}
\end{figure}

In Fig.~\ref{atomnumber} we show how $g^{(2)}(\tau)$ evolves as the number of effective atoms increases from less than one to nearly three.  For the fewest atoms [frame (a)], fluctuations in the number interacting with the cavity show up as a broad Gaussian background peak, reflecting the increase in the scattering rate when an atom is present (similar to Ref.~\cite{norris09a}.)  The beats sit on top of this background, with small visibility, a consequence of optical pumping and spontaneous emission (to modes other than the cavity mode).  The correlation function is dominated by the contribution from $g^{(2)}_{\rm max}(\tau)$ in Eqs.~(\ref{eq:fixed_speed_correlation_function}) and (\ref{eq:variable_speed_correlation_function}). As the density of the atomic beam grows [frames (b) and (c)] the background peak disappears as contributions from multiple atoms become more prominent; contributions from $|g^{(1)}_{\rm max}(\tau)|^2$, $|g^{(a)}_{\rm max}(\tau)|^2$, and $g^{(2)}_{\rm max}(\tau)$ contribute with more-or-less weight in frame (c).

\begin{figure}
\begin{center}
\includegraphics[width=0.90\linewidth]{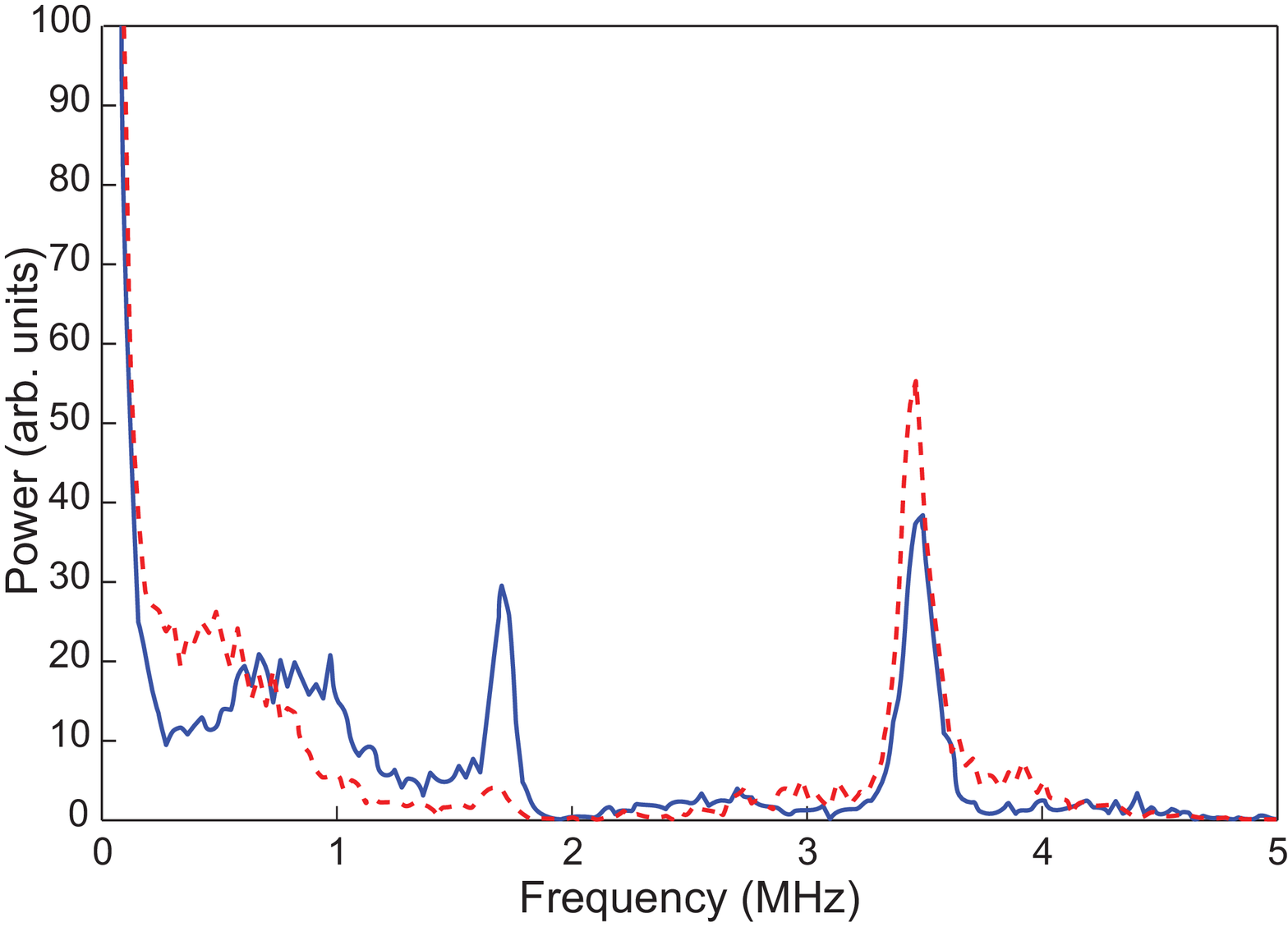}
\caption{(Color online) Sample power spectrum [FFT of the measured $g^{(2)}(\tau)$] with (red, dashed) and without (blue, solid) optical pumping of the atoms prior to entering the cavity.\label{opumping}}
\end{center}
\end{figure}

Finally, Fig.~\ref{opumping} shows the change in the spectrum when the optical pumping beam is added.  The effect on the beat frequency is minimal, revealing the robustness of the quantum interference to the initial distribution of the atoms amongst the ground-state Zeeman levels.  The biggest change is in the low-frequency components of the FFT, which correspond to atomic motion through the standing-wave mode.  This is caused by the momentum kick imparted by the optical pumping beam.

\subsection{Comparison of theory and experiment}\label{sc:exp-theo}

Outside the bad-cavity and adiabatic limit, the normalized correlation function is given by
\begin{equation}
g^{(2)}(\tau)=G^{(2)}(\tau)/I^2,
\end{equation}
with $I$ and $G^{(2)}(\tau)$ defined in Eqs.~(\ref{eq:time_averaged_intensity_again}) and (\ref{eq:quasefinal}).  For comparison with experiment, we introduce a global scale parameter, $s$, writing
\begin{eqnarray}
\label{eq:fit}
g^{(2)}(\tau)&=&1+s\left\{\vphantom{\frac{\epsilon^2}{\bar N_{\rm eff}I+\epsilon^2}}|g^{(1)}_A(\tau)|^2+|g_A^{(a)}(\tau)|^2\right.\nonumber\\
&&+\bar N_{\rm eff}^{-1}g_A^{(2)}(\tau)\nonumber\\
&&\left.+\frac{\epsilon^2}{\bar N_{\rm eff}I+\epsilon^2}2{\rm Re}[g_A^{(1)}(\tau)+g_A^{(a)}(\tau)]\right\},\mkern10mu
\end{eqnarray}
where ($\xi=1,2,a$)
\begin{equation}
g_A^{(\xi)}(\tau)=\frac{G_A^{(\xi)}(\tau)}{\bar N_{\rm eff}I+\epsilon^2}.
\end{equation}
This expression is evaluated numerically by calculating the one-atom correlation functions from the model of the atomic beam developed in Section ~\ref{sc:atomic_motion}. The speeds $v_j$ are selected from a Gaussian distribution, and the angles $\theta_j$ and $\phi_j$ from a triangular distribution.

\begin{figure}[h!]
\includegraphics[width=0.4\linewidth]{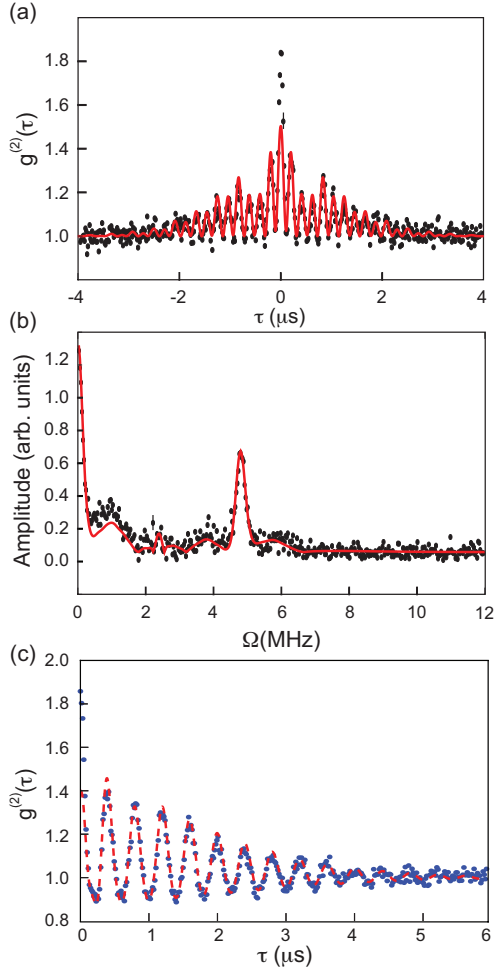}
\caption{\label{fig:comparison}
(Color online) Sample comparison of the measured (points) and calculated (solid line) correlation function, (a), and its FFT power spectrum, (b), under conditions where atomic spontaneous emission dominates the measured photon counts. A flat background is added to the calculated spectrum in order to account for residual noise in the detectors. The atomic beam parameters are: $\bar{N}_{\rm eff}=3$, $s=0.53$, $\langle v\rangle=17\mkern2mu{\rm m/s}$, $\Delta v=2\mkern2mu{\rm m/s}$, $\langle\theta\rangle=1.4\mkern4mu{\rm deg}$, $\Delta\theta=0.9\mkern4mu{\rm deg}$, $\langle\phi\rangle=0\mkern4mu{\rm deg}$, $\Delta\phi=0.7\mkern4mu{\rm deg}$. The mean number of scattered photons in the $V$ mode is $0.63$, and the birefringence background $\epsilon^2$ is 1.2\% of the driven $V$-mode photon number. (c) Sample measured (dashed line) and calculated (points) correlation functions with a small amount of coherent drive admixed in order to enhance the homodyne signal.  Parameters are: $\bar{N}_{\rm eff}=0.55$, $s=1$, $\langle v\rangle=13.5\mkern2mu{\rm m/s}$, $\Delta v=2\mkern2mu{\rm m/s}$, $\langle\theta\rangle=0.97\mkern4mu{\rm deg}$, $\Delta\theta=0.9\mkern4mu{\rm deg}$, $\langle\phi\rangle=0\mkern4mu{\rm deg}$, $\Delta\phi=0.7\mkern4mu{\rm
deg}$. The mean number of scattered photons in the $V$ mode is $1.2$, and the mixed coherent field $\epsilon^2$ is 0.5\% of the driven $V$-mode photon number, which due to the smaller number of atoms is strong enough to substantially change the signal.}
\end{figure}

Figure~\ref{fig:comparison}(a) and (b) show a fit to the experimental data and its FFT power spectrum. Experimental error bars are computed as the square root of the number of photon counts in each bin. In order to improve the fit, we adjust the mean speed and angles according to the following considerations.  The experimental data shows a low frequency modulation at short delays, the signature of a small mean inclination of the atomic trajectories away from normal to the cavity axis (see Fig.~\ref{fig:envelope}).  Comparing the FFT of the measured autocorrelation function with simulations at different mean beam angles $\langle\theta\rangle$, with $\langle\phi\rangle=0$, we find that $\langle\theta\rangle=1.4$ degrees optimizes the fit.  For small angles like this, the main parameter affecting the width of the peak at twice the Larmor frequency, around $5\mkern2mu{\rm MHz}$ in the figure, is the mean atomic speed.  A value of $\langle v\rangle=17\mkern2mu{\rm m/s}$ optimizes the fit, consistent with values expected from an LVIS \cite{lu96}.  The amplitude of the peak around $2.5\mkern2mu{\rm MHz}$ is determined by the value of $\epsilon$, which is adjusted as another free parameter.

Frame (a) of Fig.~\ref{fig:comparison} compares theory and experiment in the time domain. The reduced chi-square value for this fit is $1.3$. The major differences between the calculation and the experiment comes for delays close to zero; here the measured correlation function is
substantially larger in value compared to the calculated one. This is at least partially due to the presence of uncorrelated background light scattered into the detection path (primarily from the MOT cooling beams). Frame (b) of Fig.~\ref{fig:comparison} shows the fit in the frequency domain.  The calculated spectrum in fact goes to zero at high frequencies (larger than 7 MHz), while the experimental one remains flat due to the presence of background light in the detectors (up to about 200 MHz).  With a flat frequency background added to the calculation to account for the residual background and shot noise (as shown in the figure), the reduced chi-square value of the fit is $0.99$.  Frame (c) shows a similar correlation function after addition of a small amount of coherent drive in order to enhance the homodyne signal.  The simulation accurately captures the pronounced change in frequency and shape of the signal, with a reduced chi-square value of 1.6.

\section{Summary and Conclusions}
\label{sc:conclusions}
We have studied theoretically and experimentally how the ground-state quantum beats reported in \cite{norris10} depend on different parameters.  The fundamental beat frequency occurs at twice the Larmor frequency and is found to increase linearly with magnetic field as expected. Mixing of the driving and scattered fields produces a beat at the Larmor frequency itself.  Increasing the number of atoms brings the minimum of the oscillation to the shot noise level; this is because the many pairs of two-atom beats come to dominate the one-atom signal.  The stochastic evolution of coherence within the atomic level structure shows that the many levels of the $F=3$ to $F'=4$ transition help make the observed quantum beats robust against optical pumping.

\appendix
\section{Dipole operators}
The one atom dipole operators are given by
\begin{widetext}
\begin{eqnarray}\label{eq:dipoleoperators}
\Sigma_{\pi}&=&\sqrt{\frac{1}{4}}|g_{-3}\rangle\langle e_{-3}|+\sqrt{\frac{3}{7}}|g_{-2}\rangle\langle e_{-2}|+\sqrt{\frac{15}{28}}|g_{-1}\rangle\langle e_{-1}|+\sqrt{\frac{4}{7}}|g_{0}\rangle\langle e_{0}|+\sqrt{\frac{15}{28}}|g_{+1}\rangle\langle e_{+1}|\nonumber\\
\noalign{\vskip4pt}
&&+\sqrt{\frac{3}{7}}|g_{+2}\rangle\langle e_{+2}|+\sqrt{\frac{1}{4}}|g_{+3}\rangle\langle e_{+3}|\, ,\\
\noalign{\vskip12pt}
\Sigma_{\sigma_+}&=&\sqrt{\frac{1}{28}}|g_{-3}\rangle\langle e_{-2}|+\sqrt{\frac{3}{28}}|g_{-2}\rangle\langle e_{-1}|+\sqrt{\frac{3}{14}}|g_{-1}\rangle\langle e_{0}|+\sqrt{\frac{5}{14}}|g_{0}\rangle\langle e_{+1}|+\sqrt{\frac{15}{28}}|g_{+1}\rangle\langle e_{+2}|\nonumber\\
\noalign{\vskip4pt}
&&+\sqrt{\frac{3}{4}}|g_{+2}\rangle\langle e_{+3}|+|g_{+3}\rangle\langle e_{+4}|\, ,\\
\noalign{\vskip12pt}
\Sigma_{\sigma_-}&=&|g_{-3}\rangle\langle e_{-4}|+\sqrt{\frac{3}{4}}|g_{-2}\rangle\langle e_{-3}|+\sqrt{\frac{15}{28}}|g_{-1}\rangle\langle e_{-2}|+\sqrt{\frac{5}{14}}|g_{0}\rangle\langle
e_{-1}|+\sqrt{\frac{3}{14}}|g_{+1}\rangle\langle e_{0}|\nonumber\\
\noalign{\vskip4pt}
&&+\sqrt{\frac{3}{28}}|g_{+2}\rangle\langle e_{+1}|+\sqrt{\frac{1}{28}}|g_{+3}\rangle\langle e_{+2}|\, .
\end{eqnarray}
\end{widetext}

\acknowledgments Work supported by NSF, CONACYT, M{\'{e}}xico, and the Marsden Fund of the Royal Society of New
Zealand.


\end{document}